\documentclass[english,aps, english, A4,superscriptaddress]{revtex4-1}
\usepackage[T1]{fontenc}
\usepackage[latin9]{inputenc}
\usepackage{color}
\usepackage{babel}
\usepackage{array}
\usepackage{verbatim}
\usepackage{amsmath}
\usepackage{graphicx}
\usepackage{wasysym}
\usepackage[unicode=true,pdfusetitle,
 bookmarks=true,bookmarksnumbered=false,bookmarksopen=false,
 breaklinks=false,pdfborder={0 0 0},pdfborderstyle={},backref=false,colorlinks=true]
 {hyperref}
\hypersetup{
 linkcolor=blue, citecolor=cyan}

\makeatletter

\providecommand{\tabularnewline}{\\}

\usepackage{babel}

\makeatother

\begin{document}
\title{Lorentz transformation in Maxwell equations for slowly moving media}
\author{Xin-Li Sheng}
\email{xls@mail.ustc.edu.cn}

\affiliation{Department of Modern Physics, University of Science and Technology
of China, Hefei, Anhui 230026, China}
\author{Yang Li}
\email{leeyoung1987@ustc.edu.cn}

\affiliation{Department of Modern Physics, University of Science and Technology
of China, Hefei, Anhui 230026, China}
\author{Shi Pu}
\email{shipu@ustc.edu.cn}

\affiliation{Department of Modern Physics, University of Science and Technology
of China, Hefei, Anhui 230026, China}
\author{Qun Wang}
\email{qunwang@ustc.edu.cn}

\affiliation{Department of Modern Physics, University of Science and Technology
of China, Hefei, Anhui 230026, China}
\begin{abstract}
We use the method of field decomposition, a technique widely used
in relativistic magnetohydrodynamics, to study the small velocity
approximation (SVA) of the Lorentz transformation in Maxwell equations
for slowly moving media. The ``deformed'' Maxwell equations derived
under the SVA in the lab frame can be put into the conventional form
of Maxwell equations in the medium's comoving frame. Our results show
that the Lorentz transformation in the SVA up to $O(v/c)$ ($v$ is
the speed of the medium and $c$ is the speed of light in vacuum)
is essential to derive these equations: the time and charge density
must also change when transforming to a different frame even in the
SVA, not just the position and current density as in the Galilean
transformation. This marks the essential difference of the Lorentz
transformation from the Galilean one. We show that the integral forms
of Faraday and Ampere equations for slowly moving surfaces are consistent
with Maxwell equations. We also present Faraday equation the covariant
integral form in which the electromotive force can be defined as a
Lorentz scalar independent of the observer's frame. No evidences exist
to support an extension or modification of Maxwell equations.

\end{abstract}
\maketitle

\section{Introduction}

James Clerk Maxwell unified electricity and magnetism, the first unified
theory of physics, by constructing a set of equations now known as
Maxwell equations \citep{maxwell:1861} (for the history of Maxwell
equations, see, e.g., Ref. \citep{Rautio:2014}). Maxwell equations
are the foundation of classical physics and many technologies that
make the modern world. The Lorentz covariance is hidden in the structure
of Maxwell equations, which was first disclosed by Albert Einstein
in his well-known paper ``On the electrodynamics of moving bodies''
in 1905 that marked the discovery of special relativity \citep{Einstein:1905ve,lorentz1904,poincare1906dynamique,poincare1906note}.

Recently an extension of conventional Maxwell equations has been proposed
to charged moving media \citep{Wangzhonglin:2021} in order to describe
the power output of piezoelectric and triboelectric nanogenerators
(TENGs) \citep{Wangzhonglin:20179,Wangzhonglin:201714,Wangzhonglin:2020104272},
a new technology for fully utilizing the energy distributed in our
living environment with low quality, low amplitude and even low frequency.
The equations derived in Ref. \citep{Wangzhonglin:2021} read (in
cgs Gaussian unit and natural unit)
\begin{eqnarray}
\boldsymbol{\nabla}\cdot\mathbf{B}(t,\mathbf{x}) & = & 0\,,\nonumber \\
\boldsymbol{\nabla}\times\mathbf{E}(t,\mathbf{x}) & = & -\frac{1}{c}\left(\frac{\partial}{\partial t}+{\bf v}\cdot\boldsymbol{\nabla}\right)\mathbf{B}(t,\mathbf{x})\,,\nonumber \\
\boldsymbol{\nabla}\cdot\mathbf{D}(t,\mathbf{x}) & = & \rho_{f}(t,\mathbf{x})\,,\nonumber \\
\boldsymbol{\nabla}\times\mathbf{H}(t,\mathbf{x}) & = & \frac{1}{c}\mathbf{J}_{f}(t,\mathbf{x})+\frac{1}{c}\left(\frac{\partial}{\partial t}+{\bf v}\cdot\boldsymbol{\nabla}\right)\mathbf{D}(t,\mathbf{x})\,,\label{eq:wang-zhonglin}
\end{eqnarray}
where $\mathbf{v}$ is the velocity of the medium and assumed to be
much smaller than the speed of light $c$, and $\mathbf{D}=\mathbf{D}^{\prime}+\mathbf{P}_{s}$
with $\mathbf{D}^{\prime}$ being the conventional electric displacement
field and $\mathbf{P}_{s}$ representing the polarization owing to
the pre-existing electrostatic charges on the media induced by TENGs
\citep{Wangzhonglin:2021}. The fields $\mathbf{E}$, $\mathbf{B}$,
$\mathbf{D}^{\prime}$ and $\mathbf{H}$ are the electric, magnetic
strength, electric displacement and magnetic fields in the observer's
frame (lab frame), respectively. Note that $\mathbf{P}_{s}$ is not
linearly proportional to the electric field \citep{Wangzhonglin:2021}.
The charge conservation law in Ref. \citep{Wangzhonglin:2021} is
modified to 
\begin{equation}
\left(\frac{\partial}{\partial t}+{\bf v}\cdot\boldsymbol{\nabla}\right)\rho_{f}(t,\mathbf{x})+\boldsymbol{\nabla}\cdot\mathbf{J}_{f}(t,\mathbf{x})=0\,.\label{eq:charge-cons-wang}
\end{equation}
The differential equations in (\ref{eq:wang-zhonglin}) were derived
from an integral form of Maxwell equations \citep{Wangzhonglin:2021}.
They are different from conventional Maxwell equations in two respects:
(a) the appearance of the derivative operator $\partial/\partial t+{\bf v}\cdot\boldsymbol{\nabla}$
to replace $\partial/\partial t$; (b) the appearance of $\mathbf{P}_{s}$.
The charge conservation law is different from the conventional one
in (a).

It is obvious that the derivation of (\ref{eq:wang-zhonglin}) and
(\ref{eq:charge-cons-wang}) is not based on the Lorentz transformation
in special relativity. A natural question arises: can these equations
in (\ref{eq:wang-zhonglin}) except $\mathbf{P}_{s}$ be derived from
the Lorentz transformation under the small velocity approximation
(SVA)? The purpose of this paper is to answer this question.

In this paper, we use the (rationalized) cgs Gaussian unit \citep{Landau:1984,Jackson:1998nia}
in which electric and magnetic fields have the same unit: Gauss. In
the rationalized cgs Gaussian unit, the irrational constant $4\pi$
is absent in Maxwell equations but appears in Coulomb and Ampere force
laws among electric charges and currents respectively.  

We work in the Minkowski space-time with the metric tensor $g^{\mu\nu}=g_{\mu\nu}=\mathrm{diag}(1,-1,-1,-1)$
where $\mu,\nu=0,1,2,3$, so that we can write space-time coordinates
as $x=x^{\mu}=(x^{0},\mathbf{x})=(ct,\mathbf{x})$ and $x_{\mu}=(x_{0},-\mathbf{x})$
with $x_{0}=x^{0}=ct$. For a space position $\mathbf{x}=(\mathbf{x}_{1},\mathbf{x}_{2},\mathbf{x}_{3})$,
we do not distinguish superscripts and subscripts of its components,
$x^{i}=\mathbf{x}_{i}=\mathbf{x}^{i}$ for $i=1,2,3$. Normally we
use Greek letters to denote four-dimensional indices of four-vectors
and four-tensors, while their spatial components are denoted by space
indices (Latin letters) $i,j,k,l,m,n=1,2,3$. The four-dimensional
Levi-Civita symbols are denoted as $\epsilon^{\mu\nu\rho\sigma}$
and $\epsilon_{\mu\nu\rho\sigma}$ with the convention $\epsilon^{0123}=-\epsilon_{0123}=1$,
while the three-dimensional Levi-Civita symbol is denoted as $\epsilon_{ijk}$
with the convention $\epsilon_{123}=1$.

\section{Field decomposition and Lorentz transformation}

\label{sec:field-decomposition}In the observer's frame, the anti-symmetric
strength tensor of the electromagnetic field is given by
\begin{equation}
F^{\mu\nu}=\partial^{\mu}A^{\nu}-\partial^{\nu}A^{\mu}\,,
\end{equation}
where $x^{\mu}=(ct,\mathbf{x})$, $A^{\mu}=(A^{0},\mathbf{A})$, and
$\partial^{\mu}=(c^{-1}\partial_{t},-\boldsymbol{\nabla})$ with $\partial^{0}\equiv c^{-1}\partial_{t}\equiv c^{-1}\partial/\partial t$
and $\partial^{i}=\partial/\partial x_{i}=-\partial/\partial\mathbf{x}_{i}\equiv-\boldsymbol{\nabla}_{i}$.
The components of $F^{\mu\nu}$ are 
\begin{eqnarray}
F^{0i} & = & \partial^{0}A^{i}-\partial^{i}A^{0}=\frac{1}{c}\partial_{t}A^{i}+\nabla_{i}A^{0}=-\mathbf{E}_{i}\,,\nonumber \\
F^{ij} & = & \partial^{i}A^{j}-\partial^{j}A^{i}=-\epsilon_{ijk}\mathbf{B}_{k}\,.\label{eq:f0i-fij}
\end{eqnarray}
The components of $F_{\mu\nu}$ are then $F_{0i}=\mathbf{E}_{i}$
and $F_{ij}=-\epsilon_{ijk}\mathbf{B}_{k}$.

It is convenient to introduce a four-vector $u^{\mu}$ to decompose
$F^{\mu\nu}(x)$ into the electric and magnetic field 
\begin{equation}
F^{\mu\nu}(x)=\mathcal{E}^{\mu}(x)u^{\nu}-\mathcal{E}^{\nu}(x)u^{\mu}+\epsilon^{\mu\nu\rho\sigma}u_{\rho}\mathcal{B}_{\sigma}(x)\,,\label{eq:fmunu}
\end{equation}
where $\mathcal{E}^{\mu}$ and $\mathcal{B}^{\mu}$ are four-vectors
constructed from the electric and magnetic field respectively. Note
that $u^{\mu}$ corresponds to the four-velocity $cu^{\mu}$ and satisfies
$u_{\mu}u^{\mu}=1$, we also assume that it is a space-time constant.
They can be extracted from $F^{\mu\nu}$ by
\begin{eqnarray}
\mathcal{E}^{\mu} & = & F^{\mu\nu}u_{\nu}\,,\nonumber \\
\mathcal{B}^{\mu} & = & \frac{1}{2}\epsilon^{\mu\nu\rho\sigma}u_{\nu}F_{\rho\sigma}\equiv\widetilde{F}^{\mu\nu}u_{\nu}\,,\label{eq:extract-eb}
\end{eqnarray}
where $\widetilde{F}^{\mu\nu}=(1/2)\epsilon^{\mu\nu\alpha\beta}F_{\alpha\beta}$
is the dual of the field strength tensor. The field decomposition
(\ref{eq:fmunu}) is widely used in relativistic magnetohydrodynamics
\citep{Giacomazzo:2005jy,Huang:2011dc,Pu:2016ayh,Denicol:2018rbw}.
The Lorentz transformation of $F^{\mu\nu}$ can be realized by that
of four-vectors $\mathcal{E}^{\mu}$, $\mathcal{B}^{\mu}$ and $u^{\mu}$,
\begin{eqnarray}
F^{\prime\mu\nu}(x^{\prime}) & = & \Lambda_{\;\alpha}^{\mu}\Lambda_{\;\beta}^{\nu}F^{\alpha\beta}(x)\nonumber \\
 & = & \Lambda_{\;\alpha}^{\mu}\Lambda_{\;\beta}^{\nu}\left[\mathcal{E}^{\alpha}(x)u^{\beta}-\mathcal{E}^{\beta}(x)u^{\alpha}+\epsilon^{\alpha\beta\rho\sigma}u_{\rho}\mathcal{B}_{\sigma}(x)\right]\nonumber \\
 & = & \mathcal{E}^{\prime\mu}(x^{\prime})u^{\prime\nu}-\mathcal{E}^{\prime\nu}(x^{\prime})u^{\prime\mu}+\epsilon^{\mu\nu\rho\sigma}u_{\rho}^{\prime}\mathcal{B}_{\sigma}^{\prime}(x^{\prime})\,,\label{eq:f-f-prime}
\end{eqnarray}
where $\Lambda_{\;\alpha}^{\mu}$ denotes the Lorentz transformation
tensor and $\mathcal{E}^{\mu}(x)$ and $\mathcal{B}^{\mu}(x)$ are
transformed as four-vectors $\mathcal{E}^{\prime\mu}(x^{\prime})=\Lambda_{\;\alpha}^{\mu}\mathcal{E}^{\alpha}(x)$
and $\mathcal{B}^{\prime\mu}(x^{\prime})=\Lambda_{\;\alpha}^{\mu}\mathcal{B}^{\alpha}(x)$.
It seems that the degrees of freedom of $F^{\mu\nu}$ would be increased
because $\mathcal{E}^{\mu}$ and $\mathcal{B}^{\mu}$ are four-vectors
and would have 8 independent variables. However this is not true since
$\mathcal{E}^{\mu}$ and $\mathcal{B}^{\mu}$ are orthogonal to $u^{\mu}$,
i.e. $\mathcal{E}\cdot u=\mathcal{B}\cdot u=0$.

\begin{figure}

\caption{\label{fig:frame}The lab or observer's frame and the comoving frame
of the medium. The comoving frame moves at a three-velocity $\mathbf{v}$
relative to the lab frame. All fields and space-time in the comoving
frame are labeled with primes.}

\includegraphics[scale=0.6]{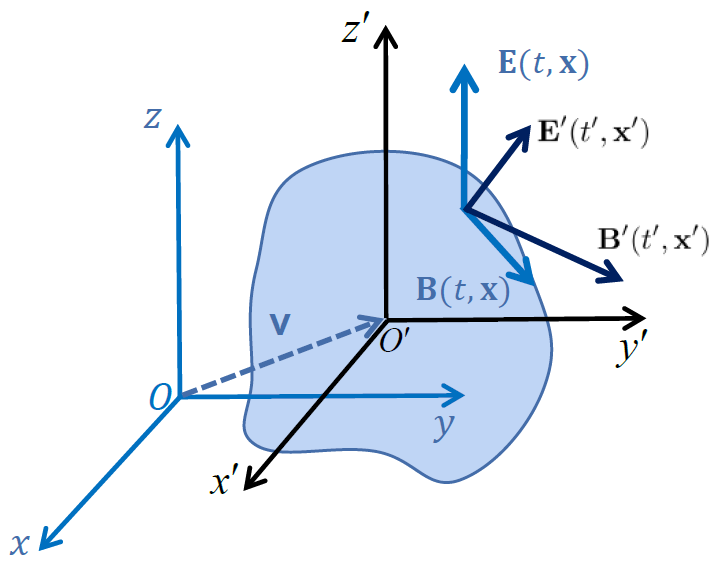}
\end{figure}

We have a freedom to choose any $u^{\mu}$ to make the decomposition
(\ref{eq:fmunu}) for $F^{\mu\nu}(x)$. As the simplest choice, we
take $u^{\mu}=u_{L}^{\mu}\equiv(1,\mathbf{0})$, which corresponds
to the lab or observer's frame as shown in Fig. \ref{fig:frame}.
Then Eq. (\ref{eq:fmunu}) has the form 
\begin{equation}
F^{\mu\nu}(x)=\mathcal{E}_{L}^{\mu}(x)u_{L}^{\nu}-\mathcal{E}_{L}^{\nu}(x)u_{L}^{\mu}+\epsilon^{\mu\nu\rho\sigma}u_{L\rho}\mathcal{B}_{L\sigma}(x)\,,\label{eq:fmunu-1}
\end{equation}
where $\mathcal{E}_{L}^{\mu}=(0,E^{1},E^{2},E^{3})=(0,\mathbf{E})$
and $\mathcal{B}_{L}^{\mu}=(0,B^{1},B^{2},B^{3})=(0,\mathbf{B})$.
The matrix form of $F^{\mu\nu}$ corresponding to $u_{L}^{\mu}$ is
then 
\begin{equation}
F^{\mu\nu}=\left(\begin{array}{cccc}
0 & -E^{1} & -E^{2} & -E^{3}\\
E^{1} & 0 & -B^{3} & B^{2}\\
E^{2} & B^{3} & 0 & -B^{1}\\
E^{3} & -B^{2} & B^{1} & 0
\end{array}\right)\,,\label{eq:em-tensor}
\end{equation}
which is just the matrix form of Eq. (\ref{eq:f0i-fij}).

As a second choice, we take $u^{\mu}=\gamma(1,\mathbf{v}/c)$ with
$\gamma=(1-v^{2}/c^{2})^{-1/2}$ being the Lorentz factor and $v\equiv|\mathbf{v}|$
being a three-velocity. In this case the electric and magnetic field
four-vectors are given by 
\begin{eqnarray}
\mathcal{E}^{\mu}(x) & = & \gamma F^{\mu0}(x)-\gamma\frac{v_{j}}{c}F^{\mu j}(x)\nonumber \\
 & = & \gamma\left(\frac{\mathbf{v}}{c}\cdot\mathbf{E},\mathbf{E}+\frac{\mathbf{v}}{c}\times\mathbf{B}\right)=\left(\frac{\mathbf{v}}{c}\cdot\boldsymbol{\mathcal{E}},\boldsymbol{\mathcal{E}}\right)\,,\nonumber \\
\mathcal{B}^{\mu}(x) & = & \frac{1}{2}\gamma\epsilon^{\mu0\rho\sigma}F_{\rho\sigma}(x)-\frac{1}{2}\gamma\frac{v_{i}}{c}\epsilon^{\mu i\rho\sigma}F_{\rho\sigma}(x)\nonumber \\
 & = & \gamma\left(\frac{\mathbf{v}}{c}\cdot\mathbf{B},\mathbf{B}-\frac{\mathbf{v}}{c}\times\mathbf{E}\right)=\left(\frac{\mathbf{v}}{c}\cdot\boldsymbol{\mathcal{B}},\boldsymbol{\mathcal{B}}\right)\,,\label{eq:eb-transform}
\end{eqnarray}
where $\mathbf{E}$, $\mathbf{B}$, $\boldsymbol{\mathcal{E}}$ and
$\boldsymbol{\mathcal{B}}$ are all functions of $x=(ct,\mathbf{x})$.
We note that $\mathcal{E}^{\mu}(x)$ and $\mathcal{B}^{\mu}(x)$ are
space-time four-vectors. We now make the Lorentz transformation for
$\mathcal{E}^{\mu}(x)$ and $\mathcal{B}^{\mu}(x)$ to the comoving
frame of the medium which moves with $\mathbf{v}$ relative to the
Lab frame (see in Fig. \ref{fig:frame}), so we have 
\begin{eqnarray}
\mathcal{E}^{\prime\mu}(x^{\prime}) & = & \Lambda_{\;\alpha}^{\mu}(\mathbf{v})\mathcal{E}^{\alpha}(x)\,,\nonumber \\
\mathcal{B}^{\prime\mu}(x^{\prime}) & = & \Lambda_{\;\alpha}^{\mu}(\mathbf{v})\mathcal{B}^{\alpha}(x)\,,\label{eq:eb-trans}
\end{eqnarray}
where $x^{\prime\mu}=\Lambda_{\;\alpha}^{\mu}(\mathbf{v})x^{\alpha}$.
With $u^{\prime\mu}=\Lambda_{\;\alpha}^{\mu}(\mathbf{v})u^{\alpha}=u_{L}^{\mu}$,
the transformation of $F^{\mu\nu}$ following Eq. (\ref{eq:f-f-prime})
reads 
\begin{equation}
F^{\prime\mu\nu}(x^{\prime})=\mathcal{E}^{\prime\mu}(x^{\prime})u_{L}^{\nu}-\mathcal{E}^{\prime\nu}(x^{\prime})u_{L}^{\mu}+\epsilon^{\mu\nu\rho\sigma}u_{L\rho}\mathcal{B}_{\sigma}^{\prime}(x^{\prime})\,.\label{eq:f-trans-u-ul}
\end{equation}
On the other hand, using $u_{L}^{\mu}$, $F^{\prime\mu\nu}(x^{\prime})$
can be rewritten as 
\begin{equation}
F^{\prime\mu\nu}(x^{\prime})=\mathcal{E}_{L}^{\prime\mu}(x^{\prime})u_{L}^{\nu}-\mathcal{E}_{L}^{\prime\nu}(x^{\prime})u_{L}^{\mu}+\epsilon^{\mu\nu\rho\sigma}u_{L\rho}\mathcal{B}_{L\sigma}^{\prime}(x^{\prime})\,.\label{eq:f-prime-ul}
\end{equation}
Comparing Eq. (\ref{eq:f-trans-u-ul}) with (\ref{eq:f-prime-ul})
we obtain 
\begin{eqnarray}
\mathcal{E}^{\prime\mu}(x^{\prime}) & = & \mathcal{E}_{L}^{\prime\mu}(x^{\prime})=(0,\mathbf{E}^{\prime}(x^{\prime}))\,,\nonumber \\
\mathcal{B}^{\prime\mu}(x^{\prime}) & = & \mathcal{B}_{L}^{\prime\mu}(x^{\prime})=(0,\mathbf{B}^{\prime}(x^{\prime}))\,,\label{eq:field-trans}
\end{eqnarray}
where $\mathbf{E}^{\prime}(x^{\prime})$ and $\mathbf{B}^{\prime}(x^{\prime})$
are the Lorentz-transformed electric and magnetic field in the moving
frame 
\begin{eqnarray}
\mathbf{E}^{\prime}(x^{\prime}) & = & \gamma\left[\mathbf{E}(x)+\frac{\mathbf{v}}{c}\times\mathbf{B}(x)\right]+(1-\gamma)\mathbf{E}_{\parallel}(x)\nonumber \\
 & = & \gamma\left[\mathbf{E}_{\perp}(x)+\frac{\mathbf{v}}{c}\times\mathbf{B}(x)\right]+\mathbf{E}_{\parallel}(x)\,,\nonumber \\
\mathbf{B}^{\prime}(x^{\prime}) & = & \gamma\left[\mathbf{B}(x)-\frac{\mathbf{v}}{c}\times\mathbf{E}(x)\right]+(1-\gamma)\mathbf{B}_{\parallel}(x)\nonumber \\
 & = & \gamma\left[\mathbf{B}_{\perp}(x)-\frac{\mathbf{v}}{c}\times\mathbf{E}(x)\right]+\mathbf{B}_{\parallel}(x)\,,\label{eq:e-b-three-dim}
\end{eqnarray}
where $\mathbf{Y}_{\parallel}=\widehat{\mathbf{v}}(\widehat{\mathbf{v}}\cdot\mathbf{Y})$
and $\mathbf{Y}_{\perp}=(1-\widehat{\mathbf{v}}\widehat{\mathbf{v}})\cdot\mathbf{Y}$
are the parallel and perpendicular part of a three-vector $\mathbf{Y}=\mathbf{E},\mathbf{B}$
to the direction $\widehat{\mathbf{v}}$ of $\mathbf{v}$. Comparing
the exact Lorentz transformation (\ref{eq:e-b-three-dim}) with $\boldsymbol{\mathcal{E}}$
and $\boldsymbol{\mathcal{B}}$ in Eq. (\ref{eq:eb-transform}), we
see the terms proportional to $1-\gamma=-[\gamma^{2}/(1+\gamma)](v^{2}/c^{2})\sim v^{2}/c^{2}$
are neglected in Eq. (\ref{eq:eb-transform}) because we only consider
the SVA up to $O(v/c)$.

\section{Maxwell equations}

The covariant form of Maxwell equations in vacuum reads 
\begin{eqnarray}
\partial_{\mu}\widetilde{F}^{\mu\nu}(x) & = & 0\,,\label{eq:homogen-maxwell}\\
\partial_{\mu}F^{\mu\nu}(x) & = & \frac{1}{c}J^{\nu}(x)\,,\label{eq:source-maxwell}
\end{eqnarray}
where $J^{\nu}=(cJ^{0},\mathbf{J})=(c\rho,\mathbf{J})$ is the four-current
density. The homogeneous equation (\ref{eq:homogen-maxwell}) gives
Faraday's law and divergence-free property of the magnetic field,
while Eq. (\ref{eq:source-maxwell}) gives Coulomb's and Ampere's
laws. So from Eqs. (\ref{eq:homogen-maxwell}) and (\ref{eq:source-maxwell}),
we obtain the conventional form of Maxwell equations in vacuum 
\begin{eqnarray}
\boldsymbol{\nabla}\cdot\mathbf{B}(x) & = & 0\,,\nonumber \\
\boldsymbol{\nabla}\times\mathbf{E}(x) & = & -\frac{1}{c}\frac{\partial\mathbf{B}(x)}{\partial t}\,,\nonumber \\
\boldsymbol{\nabla}\cdot\mathbf{E}(x) & = & \rho(x)\,,\nonumber \\
\boldsymbol{\nabla}\times\mathbf{B}(x) & = & \frac{1}{c}\mathbf{J}(x)+\frac{1}{c}\frac{\partial\mathbf{E}(x)}{\partial t}\,,\label{eq:three-dim-maxwell}
\end{eqnarray}
where all fields are functions of $x=(ct,\mathbf{x})$. The derivation
of Eq. (\ref{eq:three-dim-maxwell}) from Eqs. (\ref{eq:homogen-maxwell})
and (\ref{eq:source-maxwell}) is given in Appendix \ref{sec:derivation-dim-3}.

In the presence of medium, one can introduce the tensor $M^{\mu\nu}$
describing the polarization and magnetization of the medium. Similar
to $F^{\mu\nu}$ in Eq. (\ref{eq:fmunu}), the decomposition of $M^{\mu\nu}$
is in the following form 
\begin{equation}
M^{\mu\nu}=-(\mathcal{P}^{\mu}u^{\nu}-\mathcal{P}^{\nu}u^{\mu})+\epsilon^{\mu\nu\rho\sigma}u_{\rho}\mathcal{M}_{\sigma}(x)\,,
\end{equation}
where $\mathcal{P}^{\mu}$ and $\mathcal{M}^{\mu}$ are the polarization
and magnetization four-vector respectively. Note that there is a sign
difference between $\mathcal{P}^{\mu}$ in the above formula and $\mathcal{E}^{\mu}$
in Eq. (\ref{eq:fmunu}). Similar to Eq. (\ref{eq:extract-eb}), $\mathcal{P}^{\mu}$
and $\mathcal{M}^{\mu}$ can be extracted from $M^{\mu\nu}$ as 
\begin{eqnarray}
\mathcal{P}^{\mu} & = & -M^{\mu\nu}u_{\nu}\,,\nonumber \\
\mathcal{M}^{\mu} & = & \frac{1}{2}\epsilon^{\mu\nu\rho\sigma}u_{\nu}M_{\rho\sigma}\,.\label{eq:extract-pm}
\end{eqnarray}
Then we can define the Faraday field tensor $H^{\mu\nu}$ as 
\begin{eqnarray}
H^{\mu\nu} & = & F^{\mu\nu}-M^{\mu\nu}\nonumber \\
 & = & \mathcal{D}^{\mu}(x)u^{\nu}-\mathcal{D}^{\nu}(x)u^{\mu}+\epsilon^{\mu\nu\rho\sigma}u_{\rho}\mathcal{H}_{\sigma}(x)\,,\label{eq:faraday-tensor}
\end{eqnarray}
where $\mathcal{D}^{\mu}$ and $\mathcal{H}^{\mu}$ are the electric
displacement and magnetic field four-vector in the medium respectively
and defined by 
\begin{eqnarray}
\mathcal{D}^{\mu} & = & \mathcal{E}^{\mu}+\mathcal{P}^{\mu}\,,\nonumber \\
\mathcal{H}^{\mu} & = & \mathcal{B}^{\mu}-\mathcal{M}^{\mu}\,.
\end{eqnarray}
For homogeneous and isotropic dielectric and magnetic materials, we
have following constitutive relations \citep{minkowski1908,minkowski1910,einstein1908elektromagnetischen,pauli1981theory}
\begin{eqnarray}
\mathcal{D}^{\mu} & = & \epsilon\mathcal{E}^{\mu}\,,\nonumber \\
\mathcal{H}^{\mu} & = & \frac{1}{\mu}\mathcal{B}^{\mu}\,,\label{eq:constitutive-relations}
\end{eqnarray}
where $\epsilon$ is the electric permittivity (it is $\epsilon_{0}=1$
in vacuum) and $\mu$ is the magnetic permeability (it is $\mu_{0}=1$
in vacuum) of the medium. Note that we use cgs Gaussian unit, $\epsilon$
and $\mu$ correspond to the relative permittivity and permeability
in SI unit respectively. In terms of $F^{\mu\nu}$ and $H^{\mu\nu}$,
we have Maxwell equations in the polarized and magnetized medium
\begin{eqnarray}
\partial_{\mu}\widetilde{F}^{\mu\nu}(x) & = & 0\,,\label{eq:homogen-maxwell-1}\\
\partial_{\mu}H^{\mu\nu}(x) & = & \frac{1}{c}J_{f}^{\nu}(x)\,,\label{eq:source-maxwell-1}
\end{eqnarray}
where $J_{f}^{\mu}=(c\rho_{f},\mathbf{J}_{f})$ denotes the free four-current
density with $\rho_{f}$ and $\mathbf{J}_{f}$ being the free charge
and three-current densities. The only difference from Maxwell equations
in vacuum is the appearance of $H^{\mu\nu}$ in the equation with
the current instead of $F^{\mu\nu}$. In the presence of dielectric
and magnetic media, we can also obtain the similar equations or relations
for $\mathcal{D}^{\mu}$ and $\mathcal{H}^{\mu}$ as components of
$H^{\mu\nu}$ to Eqs. (\ref{eq:eb-transform})-(\ref{eq:e-b-three-dim})
in Sect. \ref{sec:field-decomposition}.

Corresponding to covariant Maxwell equations (\ref{eq:homogen-maxwell-1})
and (\ref{eq:source-maxwell-1}) in dielectric and magnetic media,
we have Maxwell equations in the three-dimensional form
\begin{eqnarray}
\boldsymbol{\nabla}\cdot\mathbf{B}(x) & = & 0\,,\nonumber \\
\boldsymbol{\nabla}\times\mathbf{E}(x) & = & -\frac{1}{c}\frac{\partial\mathbf{B}(x)}{\partial t}\,,\nonumber \\
\boldsymbol{\nabla}\cdot\mathbf{D}(x) & = & \rho_{f}(x)\,,\nonumber \\
\boldsymbol{\nabla}\times\mathbf{H}(x) & = & \frac{1}{c}\mathbf{J}_{f}(x)+\frac{1}{c}\frac{\partial\mathbf{D}(x)}{\partial t}\,.\label{eq:three-dim-maxwell-1}
\end{eqnarray}
The derivation of (\ref{eq:three-dim-maxwell-1}) from Eqs. (\ref{eq:homogen-maxwell-1})
and (\ref{eq:source-maxwell-1}) is similar to that of Eq. (\ref{eq:three-dim-maxwell})
in Appendix \ref{sec:derivation-dim-3}.


\section{SVA of Maxwell equations in moving frame}

We take the SVA in Eqs. (\ref{eq:eb-transform}) and (\ref{eq:e-b-three-dim})
by neglecting all $O(v^{2})$ terms which is equivalent to setting
$\gamma\approx1$, and we obtain 
\begin{eqnarray}
\boldsymbol{\mathcal{E}}(x) & \approx & {\bf E}^{\prime}(x^{\prime})\approx{\bf E}(x)+\frac{{\bf v}}{c}\times{\bf B}(x)\,,\nonumber \\
\boldsymbol{\mathcal{B}}(x) & \approx & {\bf B}^{\prime}(x^{\prime})\approx{\bf B}(x)-\frac{{\bf v}}{c}\times{\bf E}(x)\,,\label{eq:cal-e-b}
\end{eqnarray}
where $\boldsymbol{\mathcal{E}}$ and $\boldsymbol{\mathcal{B}}$
are the spatial components of $\mathcal{E}^{\mu}$ and $\mathcal{B}^{\mu}$
in (\ref{eq:eb-transform}) respectively. This indicates that $\boldsymbol{\mathcal{E}}(x)$
and $\boldsymbol{\mathcal{B}}(x)$ are the same as those used in Eq.
(2.9) in Ref. \citep{wang2022relativistic}. Similarly we also have
\begin{align}
\boldsymbol{\mathcal{D}}(x) & \approx\mathbf{D}^{\prime}(x^{\prime})\approx{\bf D}(x)+\frac{{\bf v}}{c}\times{\bf H}(x),\nonumber \\
\boldsymbol{\mathcal{H}}(x) & \approx\mathbf{H}^{\prime}(x^{\prime})\approx{\bf H}(x)-\frac{{\bf v}}{c}\times{\bf D}(x).\label{eq:cal-d-h}
\end{align}
in the presence of dielectric and magnetic media.

In order to derive Maxwell equations in terms of $\boldsymbol{\mathcal{E}}(x)$
and $\boldsymbol{\mathcal{B}}(x)$ in the SVA we can insert $F^{\mu\nu}$
in (\ref{eq:fmunu}) with $u^{\mu}=\gamma(1,\mathbf{v}/c)$ into Eqs.
(\ref{eq:homogen-maxwell}) and (\ref{eq:source-maxwell}), the covariant
Maxwell equations in vacuum. The resulting equations in three-dimensional
form read 
\begin{eqnarray}
\left(\boldsymbol{\nabla}+\frac{{\bf v}}{c^{2}}\frac{\partial}{\partial t}\right)\cdot\boldsymbol{\mathcal{B}}(x) & = & 0\,,\nonumber \\
\left(\boldsymbol{\nabla}+\frac{{\bf v}}{c^{2}}\frac{\partial}{\partial t}\right)\times\boldsymbol{\mathcal{E}}(x) & = & -\frac{1}{c}\left(\frac{\partial}{\partial t}+{\bf v}\cdot\boldsymbol{\nabla}\right)\boldsymbol{\mathcal{B}}(x)\,,\nonumber \\
\left(\boldsymbol{\nabla}+\frac{{\bf v}}{c^{2}}\frac{\partial}{\partial t}\right)\cdot\boldsymbol{\mathcal{E}}(x) & = & \rho(x)-\frac{1}{c^{2}}{\bf v}\cdot{\bf J}(x)\,,\nonumber \\
\left(\boldsymbol{\nabla}+\frac{{\bf v}}{c^{2}}\frac{\partial}{\partial t}\right)\times\boldsymbol{\mathcal{B}}(x) & = & \frac{1}{c}\left[{\bf J}(x)-\rho(x){\bf v}\right]+\frac{1}{c}\left(\frac{\partial}{\partial t}+{\bf v}\cdot\boldsymbol{\nabla}\right)\boldsymbol{\mathcal{E}}(x)\,.\label{eq:maxwell-move}
\end{eqnarray}
The derivation of above equations from Eqs. (\ref{eq:homogen-maxwell})
and (\ref{eq:source-maxwell}) is given in Appendix \ref{sec:derivation-maxwell-move}.

In the presence of homogeneous and isotropic dielectric and magnetic
materials with the constitutive relations (\ref{eq:constitutive-relations}),
we should start from Eq. (\ref{eq:source-maxwell-1}) aided by the
decomposition of $H^{\mu\nu}$ in (\ref{eq:faraday-tensor}) to obtain
non-homogeneous Maxwell equations under the SVA. The homogeneous equation
(\ref{eq:homogen-maxwell-1}) remain the same as that in vacuum and
gives the first two equations of (\ref{eq:maxwell-move}) under the
SVA. The resulting Maxwell equations for moving media now read 
\begin{eqnarray}
\left(\boldsymbol{\nabla}+\frac{{\bf v}}{c^{2}}\frac{\partial}{\partial t}\right)\cdot\boldsymbol{\mathcal{B}}(x) & = & 0\,,\nonumber \\
\left(\boldsymbol{\nabla}+\frac{{\bf v}}{c^{2}}\frac{\partial}{\partial t}\right)\times\boldsymbol{\mathcal{E}}(x) & = & -\frac{1}{c}\left(\frac{\partial}{\partial t}+{\bf v}\cdot\boldsymbol{\nabla}\right)\boldsymbol{\mathcal{B}}(x)\,,\nonumber \\
\left(\boldsymbol{\nabla}+\frac{{\bf v}}{c^{2}}\frac{\partial}{\partial t}\right)\cdot\boldsymbol{\mathcal{D}}(x) & = & \rho_{f}(x)-\frac{1}{c^{2}}{\bf v}\cdot{\bf J}_{f}(x)\,,\nonumber \\
\left(\boldsymbol{\nabla}+\frac{{\bf v}}{c^{2}}\frac{\partial}{\partial t}\right)\times\boldsymbol{\mathcal{H}}(x) & = & \frac{1}{c}\left[{\bf J}_{f}(x)-\rho_{f}(x){\bf v}\right]+\frac{1}{c}\left(\frac{\partial}{\partial t}+{\bf v}\cdot\boldsymbol{\nabla}\right)\boldsymbol{\mathcal{D}}(x)\,.\label{eq:maxwell-move-medium}
\end{eqnarray}
The derivation of above equations is similar to that of Eq. (\ref{eq:maxwell-move})
which is given in Appendix \ref{sec:derivation-maxwell-move}. Equations
in (\ref{eq:maxwell-move-medium}) are Maxwell equations in slowly
moving media seen in the lab frame. We can check the charge conservation
law by acting the operator $\boldsymbol{\nabla}+(1/c^{2}){\bf v}(\partial/\partial t)$
on the fourth equation and using the third equation of (\ref{eq:maxwell-move-medium})
as 
\begin{equation}
\left(\boldsymbol{\nabla}+\frac{{\bf v}}{c^{2}}\frac{\partial}{\partial t}\right)\cdot\left[{\bf J}_{f}(x)-\rho_{f}(x){\bf v}\right]+\left(\frac{\partial}{\partial t}+{\bf v}\cdot\boldsymbol{\nabla}\right)\left[\rho_{f}(x)-\frac{1}{c^{2}}{\bf v}\cdot{\bf J}_{f}(x)\right]=0\,,\label{eq:conservation-charge}
\end{equation}
which is equivalent to the charge conservation law in the lab frame
up to $O(v/c)$, 
\begin{equation}
\frac{\partial}{\partial t}\rho_{f}(x)+\boldsymbol{\nabla}\cdot{\bf J}_{f}(x)=0\,.
\end{equation}
Note that all terms of $O(v/c)$ cancel in Eq. (\ref{eq:conservation-charge}).
In deriving Eq. (\ref{eq:conservation-charge}) we have used the commutability
of two derivative operators 
\begin{eqnarray}
\left(\boldsymbol{\nabla}+\frac{{\bf v}}{c^{2}}\frac{\partial}{\partial t}\right)\left(\frac{\partial}{\partial t}+{\bf v}\cdot\boldsymbol{\nabla}\right) & = & \left(\frac{\partial}{\partial t}+{\bf v}\cdot\boldsymbol{\nabla}\right)\left(\boldsymbol{\nabla}+\frac{{\bf v}}{c^{2}}\frac{\partial}{\partial t}\right)\,,
\end{eqnarray}
for constant $\mathbf{v}$.

We can express ${\bf E}$ and ${\bf B}$ in terms of $\boldsymbol{\mathcal{E}}$
and $\boldsymbol{\mathcal{B}}$ using Eq. (\ref{eq:eb-transform}),
and express ${\bf D}$ and ${\bf H}$ in terms of $\boldsymbol{\mathcal{D}}$
and $\boldsymbol{\mathcal{H}}$ in a similar way. %
{} In the SVA up to $O(v/c)$, we take $\gamma\approx1$ and drop $O(v^{2}/c^{2})$
terms to obtain 
\begin{eqnarray}
{\bf E}(x) & \approx & \boldsymbol{\mathcal{E}}(x)-\frac{{\bf v}}{c}\times\boldsymbol{\mathcal{B}}(x)\,,\nonumber \\
{\bf B}(x) & \approx & \boldsymbol{\mathcal{B}}(x)+\frac{{\bf v}}{c}\times\boldsymbol{\mathcal{E}}(x)\,,\label{eq:e-b}\\
{\bf D}(x) & \approx & \boldsymbol{\mathcal{D}}(x)-\frac{{\bf v}}{c}\times\boldsymbol{\mathcal{H}}(x)\,,\nonumber \\
{\bf H}(x) & \approx & \boldsymbol{\mathcal{H}}(x)+\frac{{\bf v}}{c}\times\boldsymbol{\mathcal{D}}(x)\,.\label{eq:d-h}
\end{eqnarray}
By inserting Eqs. (\ref{eq:e-b}) and (\ref{eq:d-h}) into three-dimensional
Maxwell equations (\ref{eq:three-dim-maxwell}) and (\ref{eq:three-dim-maxwell-1})
respectively and neglecting terms of $O(v^{2}/c^{2})$, one can also
obtain Eqs. (\ref{eq:maxwell-move}) and (\ref{eq:maxwell-move-medium})
similar to the method used in Refs. \citep{wang2022relativistic,li2022comments}.

\begin{table}
\global\long\def\arraystretch{1.5}%
\begin{tabular}{>{\raggedright}m{4.5cm}|>{\raggedright}m{4.5cm}|>{\raggedright}m{4.5cm}}
\hline 
Lorentz & Lorentz {[}SVA up to $O(v/c)${]} & Galilean\tabularnewline
\hline 
\hline 
$t^{\prime}=\gamma(t-\frac{{\bf v}}{c^{2}}\cdot{\bf x})$ & $t^{\prime}\approx t-\frac{{\bf v}}{c^{2}}\cdot{\bf x}$ & $t^{\prime}=t$$\;\;\;\;\;\;\;\;\;\;\;\;(*)$\tabularnewline
\hline 
$\boldsymbol{\nabla}^{\prime}=\gamma\left(\boldsymbol{\nabla}+\frac{{\bf v}}{c^{2}}\frac{\partial}{\partial t}\right)$ & $\boldsymbol{\nabla}^{\prime}\approx\boldsymbol{\nabla}+\frac{{\bf v}}{c^{2}}\frac{\partial}{\partial t}$ & $\boldsymbol{\nabla}^{\prime}=\boldsymbol{\nabla}$$\;\;\;\;\;\;\;\;\;\;\;\;(*)$\tabularnewline
\hline 
$\rho^{\prime}(x^{\prime})=\gamma\left[\rho(x)-\frac{{\bf v}}{c^{2}}\cdot{\bf J}(x)\right]$ & $\rho^{\prime}(x^{\prime})\approx\rho(x)-\frac{{\bf v}}{c^{2}}\cdot{\bf J}(x)$ & $\rho^{\prime}(x^{\prime})=\rho(x)$$\;\;\;\;\;\;\;\;\;\;\;\;(*)$\tabularnewline
\hline 
${\bf x}^{\prime}=\gamma({\bf x}-{\bf v}t)$ & ${\bf x}^{\prime}\approx{\bf x}-{\bf v}t$ & ${\bf x}^{\prime}={\bf x}-{\bf v}t$\tabularnewline
\hline 
$\frac{\partial}{\partial t^{\prime}}=\gamma\left(\frac{\partial}{\partial t}+{\bf v}\cdot\boldsymbol{\nabla}\right)$ & $\frac{\partial}{\partial t^{\prime}}\approx\frac{\partial}{\partial t}+{\bf v}\cdot\boldsymbol{\nabla}$ & $\frac{\partial}{\partial t^{\prime}}=\frac{\partial}{\partial t}+{\bf v}\cdot\boldsymbol{\nabla}$\tabularnewline
\hline 
${\bf J}^{\prime}(x^{\prime})=\gamma\left[{\bf J}(x)-{\bf v}\rho(x)\right]$ & ${\bf J}^{\prime}(x^{\prime})\approx{\bf J}(x)-{\bf v}\rho(x)$ & ${\bf J}^{\prime}(x^{\prime})={\bf J}(x)-{\bf v}\rho(x)$\tabularnewline
\hline 
\end{tabular}\caption{\label{tab:lorentz-galilean-transf}The Lorentz transformation, its
SVA up to $O(v/c)$ and Galilean transformation for some quantities
and derivative operators. The Galilean transformation differs from
the SVA of the Lorentz transformation in the first three rows which
are labeled by \textquotedblleft$(*)$\textquotedblright . The Lorentz
transformation reduces to the Galilean one for $x^{\mu}=(ct,\mathbf{x})$
in two conditions \citep{doi:10.1121/1.1912121}: (a) $v/c\rightarrow0$;
(b) $|\mathbf{x}|\sim vt$, so does for $J^{\mu}=(c\rho,\mathbf{J})$.}
\end{table}

We can rewrite Eq. (\ref{eq:maxwell-move-medium}) in a compact form
if we replace $\boldsymbol{\mathcal{E}}(x)$, $\boldsymbol{\mathcal{B}}(x)$,
$\boldsymbol{\mathcal{D}}(x)$ and $\boldsymbol{\mathcal{H}}(x)$
by ${\bf E}^{\prime}(x^{\prime})$, ${\bf B}^{\prime}(x^{\prime})$,
${\bf D}^{\prime}(x^{\prime})$ and ${\bf H}^{\prime}(x^{\prime})$
following Eqs. (\ref{eq:cal-e-b}) and (\ref{eq:cal-d-h}). The resulting
equations read
\begin{eqnarray}
\boldsymbol{\nabla}^{\prime}\cdot{\bf B}^{\prime}(x^{\prime}) & = & 0\,,\nonumber \\
\boldsymbol{\nabla}^{\prime}\times{\bf E}^{\prime}(x^{\prime}) & = & -\frac{1}{c}\frac{\partial}{\partial t^{\prime}}{\bf B}^{\prime}(x^{\prime})\,,\nonumber \\
\boldsymbol{\nabla}^{\prime}\cdot{\bf D}^{\prime}(x^{\prime}) & = & \rho_{f}^{\prime}(x^{\prime})\,,\nonumber \\
\boldsymbol{\nabla}^{\prime}\times{\bf H}^{\prime}(x^{\prime}) & = & \frac{1}{c}{\bf J}_{f}^{\prime}(x^{\prime})+\frac{1}{c}\frac{\partial}{\partial t^{\prime}}{\bf D}^{\prime}(x^{\prime})\,,\label{eq:maxwell-move-medium-1}
\end{eqnarray}
where we have used the Lorentz transformation in the SVA up to $O(v/c)$
for quantities and operators listed in the second column of Table
\ref{tab:lorentz-galilean-transf}. Also we can rewrite the charge
conservation law (\ref{eq:conservation-charge}) in terms of quantities
in the comoving frame 
\begin{equation}
\frac{\partial}{\partial t^{\prime}}\rho_{f}^{\prime}(x^{\prime})+\boldsymbol{\nabla}^{\prime}\cdot{\bf J}_{f}^{\prime}(x^{\prime})=0\,,\label{eq:charge-cons-comov}
\end{equation}
which can be proved by taking a divergence $\boldsymbol{\nabla}^{\prime}$
of the fourth equation and using the third equation of (\ref{eq:maxwell-move-medium-1}).
We see in Table \ref{tab:lorentz-galilean-transf} that the Lorentz
transformation in the SVA obviously differs from the Galilean transformation
in the first three rows: the time, the charge density and the space-derivative
operator $\boldsymbol{\nabla}$ are not invariant in the former, while
they are invariant in the latter. However, different from the cases
of the space-time and charge-current density, the Galilean transformation
of electric and magnetic fields is not well-defined, see, e.g., Refs.
\citep{bellac:1973,daixi:2022,daixi:2022en} for discussions of this
topic.

Equation (\ref{eq:maxwell-move-medium-1}) is nothing but Maxwell
equations in the comoving frame of the medium. It is not surprising
that Maxwell equations have the same form in the comoving frame as
shown in (\ref{eq:maxwell-move-medium-1}). However, what makes Eq.
(\ref{eq:maxwell-move-medium}) {[}another form of (\ref{eq:maxwell-move-medium-1}){]}
special is that all fields are in the comoving frame while the space-time
coordinates are in the lab frame. The physical meaning of Eq. (\ref{eq:maxwell-move-medium})
needs to clarified especially when applied to real problems such as
TENGs.

We see that Eqs. (\ref{eq:maxwell-move-medium}) and (\ref{eq:conservation-charge})
look similar to Eqs. (\ref{eq:wang-zhonglin}) and (\ref{eq:charge-cons-wang})
derived in Ref. \citep{Wangzhonglin:2021}. But the main difference
lies in that all fields (including charge and current densities) in
Eqs. (\ref{eq:maxwell-move-medium}) and (\ref{eq:conservation-charge})
are those in the comoving frame, while all fields in Eqs. (\ref{eq:wang-zhonglin})
and (\ref{eq:charge-cons-wang}) are those in the lab frame. Another
difference is that $\boldsymbol{\nabla}^{\prime}=\boldsymbol{\nabla}+({\bf v}/c^{2})\partial_{t}$
appears in Eqs. (\ref{eq:maxwell-move-medium}) and (\ref{eq:conservation-charge})
instead of $\boldsymbol{\nabla}$ in Eqs. (\ref{eq:wang-zhonglin})
and (\ref{eq:charge-cons-wang}). These differences seems to indicate
that Eq. (\ref{eq:wang-zhonglin}) might be related to the Galilean
transformation instead of the SVA of the Lorentz one. Also, the electric
and magnetic fields are thought to move with the medium from the arguments
of Ref. \citep{wangqing:2022}, which behave like scalar fields.

The conditions that $\boldsymbol{\nabla}^{\prime}=\boldsymbol{\nabla}+({\bf v}/c^{2})\partial_{t}$
can be approximated as $\boldsymbol{\nabla}$ are 
\begin{eqnarray}
\frac{1}{c^{2}}\frac{\partial}{\partial t}({\bf v}\cdot\boldsymbol{\mathcal{F}}) & \ll & \boldsymbol{\nabla}\cdot\boldsymbol{\mathcal{F}},\;\;\mathrm{for}\;\boldsymbol{\mathcal{F}}=\boldsymbol{\mathcal{D}},\boldsymbol{\mathcal{B}},{\bf J}_{f}-\rho_{f}{\bf v}\nonumber \\
\frac{1}{c^{2}}\frac{\partial}{\partial t}({\bf v}\times\boldsymbol{\mathcal{F}}) & \ll & \boldsymbol{\nabla}\times\boldsymbol{\mathcal{F}}.\;\;\mathrm{for}\;\boldsymbol{\mathcal{F}}=\boldsymbol{\mathcal{E}},\boldsymbol{\mathcal{H}}\label{eq:lor-ga-cond}
\end{eqnarray}
In the space of the wave number $k$ and the frequency $\omega$ of
above fields, the above conditions can be put into a general form
\begin{equation}
\frac{\omega}{k}\ll\frac{c^{2}}{v}.\label{eq:condition-1}
\end{equation}
Note that in the SVA of Lorentz transformation we have $v\ll c$ which
leads to $c^{2}/v\gg c$.

The conditions for some four-vectors such as $x^{\mu}=(ct,\mathbf{x})$
and $J^{\mu}=(c\rho,\mathbf{J})$ {[}or $J_{f}^{\mu}=(c\rho_{f},\mathbf{J}_{f})${]}
that the Lorentz transformation reduces to the Galilean one are 
\begin{equation}
|\mathbf{x}|\sim vt\ll ct,\;\;\;\left|\mathbf{J}\right|\sim\rho v\ll\rho c.\label{eq:condition-2}
\end{equation}
So that we have $t^{\prime}\approx t$ and $\rho^{\prime}(x^{\prime})\approx\rho(x)$
up to $O(v/c)$. However, the Galilean transformation for electric
and magnetic fields are not well-defined \citep{bellac:1973,daixi:2022en}.
There are two limits in applications: the electric quasi-static limit
in which the system is dominated by $\rho$ and $\mathbf{E}$ relative
to $\mathbf{J}$ and $\mathbf{B}$ respectively, and the magnetic
quasi-static limit in which the system is dominated by $\mathbf{J}$
and $\mathbf{B}$ relative to $\rho$ and $\mathbf{E}$ respectively.
It can be checked if the conditions (\ref{eq:lor-ga-cond})-(\ref{eq:condition-2})
as well as above two limits are really satisfied in TENGs.

Let us comment on the main results, Eqs. (V.7) and (V.8), of Ref.
\citep{li2022comments}. These equations mix fields of different frames
and were previously derived by Pauli \citep{pauli1981theory}. The
fields $\mathbf{E}^{*}(x)$ and $\mathbf{H}^{*}(x)$ defined by Pauli
are actually $\boldsymbol{\mathcal{E}}(x)$ and $\boldsymbol{\mathcal{H}}(x)$
in the SVA, 
\begin{eqnarray}
\mathbf{E}^{*}(x) & \equiv & {\bf E}(x)+\frac{{\bf v}}{c}\times{\bf B}(x)\approx\boldsymbol{\mathcal{E}}(x)\approx{\bf E}^{\prime}(x^{\prime}),\nonumber \\
\mathbf{H}^{*}(x) & \equiv & {\bf H}(x)-\frac{{\bf v}}{c}\times{\bf D}(x)\approx\boldsymbol{\mathcal{H}}(x)\approx{\bf H}^{\prime}(x^{\prime}),
\end{eqnarray}
Then one can verify Eq. (274) of Ref. \citep{pauli1981theory}, 
\begin{eqnarray}
\boldsymbol{\nabla}\times\boldsymbol{\mathcal{E}}(x) & = & \boldsymbol{\nabla}\times{\bf E}(x)+\frac{1}{c}\boldsymbol{\nabla}\times\left[{\bf v}\times{\bf B}(x)\right]\nonumber \\
 & = & -\frac{1}{c}\left(\frac{\partial}{\partial t}+{\bf v}\cdot\boldsymbol{\nabla}\right)\mathbf{B}(x),\label{eq:pauli-1}\\
\boldsymbol{\nabla}\times\boldsymbol{\mathcal{H}}(x) & = & \boldsymbol{\nabla}\times{\bf H}(x)-\frac{1}{c}\boldsymbol{\nabla}\times\left[{\bf v}\times{\bf D}(x)\right]\nonumber \\
 & = & \frac{1}{c}\mathbf{J}_{f}(x)+\frac{1}{c}\frac{\partial{\bf D}(x)}{\partial t}-\frac{{\bf v}}{c}\boldsymbol{\nabla}\cdot{\bf D}(x)+\frac{{\bf v}}{c}\cdot\boldsymbol{\nabla}{\bf D}(x)\nonumber \\
 & = & \frac{1}{c}\left[\mathbf{J}_{f}(x)-\rho_{f}(x){\bf v}\right]+\frac{1}{c}\left(\frac{\partial}{\partial t}+{\bf v}\cdot\boldsymbol{\nabla}\right)\mathbf{D}(x)\label{eq:pauli-2}
\end{eqnarray}
where we have used Maxwell equations in (\ref{eq:three-dim-maxwell-1}).
Note that $\mathbf{J}_{f}(x)-\rho_{f}(x){\bf v}$ in Eq. (\ref{eq:pauli-2})
can be approximated as $\mathbf{J}_{f}^{\prime}(x^{\prime})$ in the
SVA of Lorentz transformation or Galilean transformation, see Table
\ref{tab:lorentz-galilean-transf}. In the same spirit we can rewrite
the charge conservation equation as 
\begin{equation}
\left(\frac{\partial}{\partial t}+{\bf v}\cdot\boldsymbol{\nabla}\right)\rho_{f}(t,\mathbf{x})+\boldsymbol{\nabla}\cdot\left[\mathbf{J}_{f}(x)-\rho_{f}(x){\bf v}\right]=0.\label{eq:charge-cons-mix}
\end{equation}
One can verify that Eq. (\ref{eq:pauli-1}) is equivalent to the second
equation of (\ref{eq:maxwell-move-medium}) and Eq. (\ref{eq:pauli-2})
is equivalent to the fourth equation of (\ref{eq:maxwell-move-medium})
after expressing $\mathbf{B}(x)$ in terms of $\boldsymbol{\mathcal{E}}(x)$
and $\boldsymbol{\mathcal{B}}(x)$ following Eq. (\ref{eq:e-b}) and
$\mathbf{D}(x)$ in terms of $\boldsymbol{\mathcal{D}}(x)$ and $\boldsymbol{\mathcal{H}}(x)$
following Eq. (\ref{eq:d-h}). We classify Eqs. (\ref{eq:pauli-1})-(\ref{eq:charge-cons-mix})
to Maxwell equations in case (d) in Table \ref{tab:ME-form}, and
we will show in Sect. \ref{sec:integral-form} that these equations
are actually Faraday and Ampere equations for moving surfaces. Note
that Eqs. (\ref{eq:pauli-1})-(\ref{eq:charge-cons-mix}) are also
different from Eqs. (\ref{eq:wang-zhonglin}) and (\ref{eq:charge-cons-wang}).

In Table \ref{tab:ME-form}, we also list other three equivalent forms
of Maxwell equations (of course there are many other equivalent forms
besides those listed in the table).

\begin{table}
\caption{\label{tab:ME-form}Maxwell and charge conservation equations in different
forms which are all equivalent in the SVA of Lorentz transformation
up to $O(v/c)$. These are fields in the lab frame: $\mathbf{E}(x)$,
$\mathbf{B}(x)$, $\mathbf{D}(x)$, $\mathbf{H}(x)$, $\rho_{f}(x)$
and $\mathbf{J}_{f}(x)$. These are fields in the comoving frame:
${\bf E}^{\prime}(x^{\prime})$, ${\bf B}^{\prime}(x^{\prime})$,
${\bf D}^{\prime}(x^{\prime})$, ${\bf H}^{\prime}(x^{\prime})$,
$\rho_{f}^{\prime}(x^{\prime})$ and ${\bf J}_{f}^{\prime}(x^{\prime})$.
Note that $\boldsymbol{\mathcal{E}}(x)$ is approximately ${\bf E}^{\prime}(x^{\prime})$
but expressed in the lab-frame space-time since it is a linear combination
of $\mathbf{E}(x)$ and $\mathbf{B}(x)$, so do other fields in calligraphic
fonts. We use the (rationalized) cgs Gaussian unit in which electric
and magnetic fields have the same unit: Gauss.}

\global\long\def\arraystretch{1.8}%
\begin{tabular}{|l|c|}
\hline 
\multicolumn{2}{|l|}{Transformation of fields}\tabularnewline
\hline 
$\boldsymbol{\mathcal{E}}(x)\approx{\bf E}^{\prime}(x^{\prime})\approx{\bf E}(x)+\frac{{\bf v}}{c}\times{\bf B}(x)$ & ${\bf J}_{f}^{\prime}(x^{\prime})\approx{\bf J}_{f}(x)-\rho_{f}(x){\bf v}$\tabularnewline
\hline 
$\boldsymbol{\mathcal{B}}(x)\approx{\bf B}^{\prime}(x^{\prime})\approx{\bf B}(x)-\frac{{\bf v}}{c}\times{\bf E}(x)$ & $\rho_{f}^{\prime}(x^{\prime})\approx\rho_{f}(x)-\frac{{\bf v}}{c^{2}}\cdot{\bf J}_{f}(x)$\tabularnewline
\hline 
$\boldsymbol{\mathcal{D}}(x)\approx{\bf D}^{\prime}(x^{\prime})\approx{\bf D}(x)+\frac{{\bf v}}{c}\times{\bf H}(x)$ & $t^{\prime}\approx t-\frac{{\bf v}}{c^{2}}\cdot{\bf x}$, ${\bf x}^{\prime}\approx{\bf x}-{\bf v}t$\tabularnewline
\hline 
$\boldsymbol{\mathcal{H}}(x)\approx{\bf H}^{\prime}(x^{\prime})\approx{\bf H}(x)-\frac{{\bf v}}{c}\times{\bf D}(x)$ & $\frac{\partial}{\partial t^{\prime}}\approx\frac{\partial}{\partial t}+{\bf v}\cdot\boldsymbol{\nabla}$,
$\boldsymbol{\nabla}^{\prime}\approx\boldsymbol{\nabla}+\frac{{\bf v}}{c^{2}}\frac{\partial}{\partial t}$\tabularnewline
\hline 
\end{tabular}\\\vspace{0.5cm}%
\begin{tabular}{|l|l|}
\hline 
(a) Lab frame & (b) Comoving frame\tabularnewline
\hline 
$\boldsymbol{\nabla}\cdot\mathbf{B}(x)=0$ & $\boldsymbol{\nabla}^{\prime}\cdot{\bf B}^{\prime}(x^{\prime})=0$\tabularnewline
$\boldsymbol{\nabla}\times\mathbf{E}(x)=-\frac{1}{c}\frac{\partial\mathbf{B}(x)}{\partial t}$ & $\boldsymbol{\nabla}^{\prime}\times{\bf E}^{\prime}(x^{\prime})=-\frac{1}{c}\frac{\partial}{\partial t^{\prime}}{\bf B}^{\prime}(x^{\prime})$\tabularnewline
$\boldsymbol{\nabla}\cdot\mathbf{D}(x)=\rho_{f}(x)$ & $\boldsymbol{\nabla}^{\prime}\cdot{\bf D}^{\prime}(x^{\prime})=\rho_{f}^{\prime}(x^{\prime})$\tabularnewline
$\boldsymbol{\nabla}\times\mathbf{H}(x)=\frac{1}{c}\mathbf{J}_{f}(x)+\frac{1}{c}\frac{\partial\mathbf{D}(x)}{\partial t}$ & $\boldsymbol{\nabla}^{\prime}\times{\bf H}^{\prime}(x^{\prime})=\frac{1}{c}{\bf J}_{f}^{\prime}(x^{\prime})+\frac{1}{c}\frac{\partial}{\partial t^{\prime}}{\bf D}^{\prime}(x^{\prime})$\tabularnewline
\hline 
$\frac{\partial}{\partial t}\rho_{f}(x)+\boldsymbol{\nabla}\cdot{\bf J}_{f}(x)=0$ & $\frac{\partial}{\partial t^{\prime}}\rho_{f}^{\prime}(x^{\prime})+\boldsymbol{\nabla}^{\prime}\cdot{\bf J}_{f}^{\prime}(x^{\prime})=0$\tabularnewline
\hline 
\end{tabular}\\\vspace{0.5cm}%
\begin{tabular}{|l|}
\hline 
(c) Fields in the comoving frame and space-time in the lab frame\tabularnewline
\hline 
$\left(\boldsymbol{\nabla}+\frac{{\bf v}}{c^{2}}\frac{\partial}{\partial t}\right)\cdot\boldsymbol{\mathcal{B}}(x)=0$\tabularnewline
$\left(\boldsymbol{\nabla}+\frac{{\bf v}}{c^{2}}\frac{\partial}{\partial t}\right)\times\boldsymbol{\mathcal{E}}(x)=-\frac{1}{c}\left(\frac{\partial}{\partial t}+{\bf v}\cdot\boldsymbol{\nabla}\right)\boldsymbol{\mathcal{B}}(x)$\tabularnewline
$\left(\boldsymbol{\nabla}+\frac{{\bf v}}{c^{2}}\frac{\partial}{\partial t}\right)\cdot\boldsymbol{\mathcal{D}}(x)=\rho_{f}(x)-\frac{{\bf v}}{c^{2}}\cdot{\bf J}_{f}(x)$\tabularnewline
$\left(\boldsymbol{\nabla}+\frac{{\bf v}}{c^{2}}\frac{\partial}{\partial t}\right)\times\boldsymbol{\mathcal{H}}(x)=\frac{1}{c}\left[{\bf J}_{f}(x)-\rho_{f}(x){\bf v}\right]+\frac{1}{c}\left(\frac{\partial}{\partial t}+{\bf v}\cdot\boldsymbol{\nabla}\right)\boldsymbol{\mathcal{D}}(x)$\tabularnewline
\hline 
$\left(\frac{\partial}{\partial t}+{\bf v}\cdot\boldsymbol{\nabla}\right)\left[\rho_{f}(x)-\frac{1}{c^{2}}{\bf v}\cdot{\bf J}_{f}(x)\right]+\left(\boldsymbol{\nabla}+\frac{{\bf v}}{c^{2}}\frac{\partial}{\partial t}\right)\cdot\left[{\bf J}_{f}(x)-\rho_{f}(x){\bf v}\right]=0$\tabularnewline
\hline 
\end{tabular}\\\vspace{0.5cm}%
\begin{tabular}{|l|}
\hline 
(d) Fields in both frames and space-time in the lab frame\tabularnewline
\hline 
$\boldsymbol{\nabla}\cdot\mathbf{B}(x)=0$\tabularnewline
$\boldsymbol{\nabla}\times\boldsymbol{\mathcal{E}}(x)=-\frac{1}{c}\left(\frac{\partial}{\partial t}+{\bf v}\cdot\boldsymbol{\nabla}\right)\mathbf{B}(x)$\tabularnewline
$\boldsymbol{\nabla}\cdot\mathbf{D}(x)=\rho_{f}(x)$\tabularnewline
$\boldsymbol{\nabla}\times\boldsymbol{\mathcal{H}}(x)=\frac{1}{c}\left[{\bf J}_{f}(x)-\rho_{f}(x){\bf v}\right]+\frac{1}{c}\left(\frac{\partial}{\partial t}+{\bf v}\cdot\boldsymbol{\nabla}\right)\mathbf{D}(x)$\tabularnewline
\hline 
$\left(\frac{\partial}{\partial t}+{\bf v}\cdot\boldsymbol{\nabla}\right)\rho_{f}(x)+\boldsymbol{\nabla}\cdot\left[\mathbf{J}_{f}(x)-\rho_{f}(x){\bf v}\right]=0$\tabularnewline
\hline 
\end{tabular}\\
\end{table}


\section{Discussions about extended Hertz equations and constitutive relations}

In order to derive the extended Hertz equations for $\boldsymbol{\mathcal{E}}(x)$
and $\boldsymbol{\mathcal{B}}(x)$ in moving media with homogeneous
and isotropic dielectric and magnetic properties, we need to express
$\boldsymbol{\mathcal{D}}(x)$ and $\boldsymbol{\mathcal{H}}(x)$
in the fourth equation of (\ref{eq:maxwell-move-medium}) in terms
of $\boldsymbol{\mathcal{E}}(x)$ and $\boldsymbol{\mathcal{B}}(x)$
using the covariant linear constitutive relations 
\begin{eqnarray}
\boldsymbol{\mathcal{D}}(x) & = & \epsilon\boldsymbol{\mathcal{E}}(x),\nonumber \\
\boldsymbol{\mathcal{H}}(x) & = & \frac{1}{\mu}\boldsymbol{\mathcal{B}}(x),\label{eq:dim-3-constitutive}
\end{eqnarray}
following Eq. (\ref{eq:constitutive-relations}). The above constitutive
relations lead to the ones in fields of the lab frame up to $O(v/c)$
\begin{eqnarray}
\mathbf{D}(x) & = & \epsilon\mathbf{E}(x)+\frac{\alpha}{\widetilde{c}^{2}}\frac{\mathbf{v}}{c}\times\mathbf{H}(x),\nonumber \\
\mathbf{B}(x) & = & \mu\mathbf{H}(x)-\frac{\alpha}{\widetilde{c}^{2}}\frac{\mathbf{v}}{c}\times\mathbf{E}(x),
\end{eqnarray}
where $\widetilde{c}\equiv1/\sqrt{\epsilon\mu}$ is the speed of light
in the medium and $\alpha\equiv1-\widetilde{c}^{2}$ is a constant
related to the medium and it is vanishing in vacuum. Using (\ref{eq:dim-3-constitutive}),
the second and fourth equations of (\ref{eq:maxwell-move-medium})
give 
\begin{eqnarray}
\boldsymbol{\nabla}\times\boldsymbol{\mathcal{E}}(x) & = & -\frac{1}{c}\left(\frac{\partial}{\partial t}+\alpha{\bf v}\cdot\boldsymbol{\nabla}\right)\boldsymbol{\mathcal{B}}(x)+\frac{1}{\epsilon c^{2}}{\bf v}\times{\bf J}_{f}(x)-\frac{\widetilde{c}^{2}}{c}\boldsymbol{\nabla}\left[{\bf v}\cdot\boldsymbol{\mathcal{B}}(x)\right]\nonumber \\
 & \approx & -\frac{1}{c}\frac{\partial}{\partial t}\boldsymbol{\mathcal{B}}(x)+\frac{1}{c}\boldsymbol{\nabla}\times\left[\alpha\mathbf{v}\times\boldsymbol{\mathcal{B}}(x)\right]+\frac{1}{\epsilon c^{2}}{\bf v}\times{\bf J}_{f}(x)-\frac{\widetilde{c}^{2}}{c}\boldsymbol{\nabla}\left[{\bf v}\cdot\boldsymbol{\mathcal{B}}(x)\right]\,,\nonumber \\
\boldsymbol{\nabla}\times\boldsymbol{\mathcal{B}}(x) & = & \frac{1}{\widetilde{c}^{2}c}\left(\frac{\partial}{\partial t}+\alpha{\bf v}\cdot\boldsymbol{\nabla}\right)\boldsymbol{\mathcal{E}}(x)+\frac{\mu}{c}\left[{\bf J}_{f}(x)-\rho_{f}(x)\mathbf{v}\right]+\frac{1}{c}\boldsymbol{\nabla}\left[{\bf v}\cdot\boldsymbol{\mathcal{E}}(x)\right]\nonumber \\
 & \approx & \frac{1}{\widetilde{c}^{2}c}\frac{\partial}{\partial t}\boldsymbol{\mathcal{E}}(x)-\frac{1}{\widetilde{c}^{2}c}\boldsymbol{\nabla}\times\left[\alpha\mathbf{v}\times\boldsymbol{\mathcal{E}}(x)\right]\nonumber \\
 &  & +\mu\frac{1}{c}\left[{\bf J}_{f}(x)-\widetilde{c}^{2}\rho_{f}(x)\mathbf{v}\right]+\frac{1}{c}\boldsymbol{\nabla}\left[{\bf v}\cdot\boldsymbol{\mathcal{E}}(x)\right]\,,\label{eq:hertz-3}
\end{eqnarray}
where we have expressed $\partial\boldsymbol{\mathcal{E}}(x)/\partial t$
and $\partial\boldsymbol{\mathcal{B}}(x)/\partial t$ in the second
and fourth equation of (\ref{eq:maxwell-move-medium}) in terms of
$\boldsymbol{\mathcal{B}}$ and $\boldsymbol{\mathcal{E}}$ respectively
by using the other equation. We see the modified derivative time operators
in medium in two equations have the same form, $\widetilde{\partial}_{t^{\prime}}\equiv\partial_{t}+\alpha{\bf v}\cdot\boldsymbol{\nabla}$.
Equation (\ref{eq:hertz-3}) can be rewritten in terms of $\mathbf{E}(x)$
and $\mathbf{B}(x)$ using Eqs. (\ref{eq:cal-e-b}) {[}and the same
relations for $\boldsymbol{\mathcal{D}}$ and $\boldsymbol{\mathcal{H}}$
to $\mathbf{D}$ and $\mathbf{H}${]} and (\ref{eq:dim-3-constitutive})
as 
\begin{eqnarray}
\boldsymbol{\nabla}\times({\bf E}+\frac{\alpha}{c}{\bf v}\times{\bf B}) & = & -\frac{1}{c}\frac{\partial{\bf B}}{\partial t}+\frac{\alpha}{c}\boldsymbol{\nabla}\times({\bf v}\times{\bf B})\,,\nonumber \\
\boldsymbol{\nabla}\times({\bf H}-\frac{\alpha}{c}\alpha{\bf v}\times{\bf D}) & = & \frac{1}{c}{\bf J}_{f}(x)+\frac{1}{c}\frac{\partial{\bf D}}{\partial t}-\frac{\alpha}{c}\boldsymbol{\nabla}\times(\mathbf{v}\times\mathbf{D})\,,
\end{eqnarray}
which is consistent with the corresponding equations in Refs. \citep{Rozov+2015+1019+1024,li2022comments}.
If we neglect ${\bf v}\cdot\boldsymbol{\mathcal{B}}$ and ${\bf v}\cdot\boldsymbol{\mathcal{E}}$
terms in Eq. (\ref{eq:hertz-3}) and calculate the dispersion relation
without free charges and currents, we obtain two modes: one mode has
the group velocity less than $\widetilde{c}$, while the other mode
has the group velocity larger than $\widetilde{c}$ and then is superluminal.
These modes are observed in the lab frame so the dispersion relations
depend on the velocity $\mathbf{v}$ of the medium. However, if we
work in the comoving frame of the medium with Eq. (\ref{eq:maxwell-move-medium-1}),
we will see that all modes propagate at the speed of light $\widetilde{c}$
without any dispersion.

We note that in deriving Eq. (\ref{eq:hertz-3}), we have used the
covariant constitutive relations in (\ref{eq:dim-3-constitutive})
for the fields in the comoving frame. If one uses the constitutive
relations for the fields in the lab frame 
\begin{eqnarray}
{\bf D}(x) & = & \epsilon\mathbf{E}(x)\,,\nonumber \\
{\bf H}(x) & = & \frac{1}{\mu}{\bf B}(x)\,,\label{eq:constitutive-lab}
\end{eqnarray}
which are only valid for static media but not for moving media, one
would obtain up to $O(v/c)$
\begin{eqnarray}
\boldsymbol{\nabla}\times\boldsymbol{\mathcal{E}}(x) & = & -\frac{1}{c}\left(\frac{\partial}{\partial t}+\alpha{\bf v}\cdot\boldsymbol{\nabla}\right)\boldsymbol{\mathcal{B}}(x)-\frac{\widetilde{c}^{2}}{c}\boldsymbol{\nabla}\left[{\bf v}\cdot\boldsymbol{\mathcal{B}}(x)\right]\nonumber \\
 & \approx & -\frac{1}{c}\frac{\partial}{\partial t}\boldsymbol{\mathcal{B}}(x)+\frac{1}{c}\boldsymbol{\nabla}\times\left[\alpha\mathbf{v}\times\boldsymbol{\mathcal{B}}(x)\right]-\frac{\widetilde{c}^{2}}{c}\boldsymbol{\nabla}\left[{\bf v}\cdot\boldsymbol{\mathcal{B}}(x)\right]\,,\nonumber \\
\boldsymbol{\nabla}\times\boldsymbol{\mathcal{B}}(x) & = & \frac{1}{\widetilde{c}^{2}c}\left(\frac{\partial}{\partial t}-\alpha{\bf v}\cdot\boldsymbol{\nabla}\right)\boldsymbol{\mathcal{E}}(x)+\frac{1}{\widetilde{c}^{2}c}\boldsymbol{\nabla}\left[{\bf v}\cdot\boldsymbol{\mathcal{E}}(x)\right]\nonumber \\
 & \approx & \frac{1}{\widetilde{c}^{2}c}\frac{\partial}{\partial t}\boldsymbol{\mathcal{E}}(x)+\frac{1}{\widetilde{c}^{2}c}\boldsymbol{\nabla}\times\left[\alpha\mathbf{v}\times\boldsymbol{\mathcal{E}}(x)\right]+\frac{1}{\widetilde{c}^{2}c}\boldsymbol{\nabla}\left[{\bf v}\cdot\boldsymbol{\mathcal{E}}(x)\right]\,,\label{eq:hertz-2}
\end{eqnarray}
where the charge and current densities have been neglected. Note about
the opposite sign of $\alpha$ terms in modified derivative time operators
$\widetilde{\partial}_{t^{\prime}}\equiv\partial_{t}\pm\alpha{\bf v}\cdot\boldsymbol{\nabla}$
in medium, which clearly indicates that the Lorentz covariance is
lost in the moving medium. The similar equations are derived in Ref.
\citep{wang2022relativistic} except ${\bf v}\cdot\boldsymbol{\mathcal{B}}$
and ${\bf v}\cdot\boldsymbol{\mathcal{E}}$ terms. The opposite sign
of $\alpha$ terms leads to the superluminal problem (without ${\bf v}\cdot\boldsymbol{\mathcal{B}}$
and ${\bf v}\cdot\boldsymbol{\mathcal{E}}$ terms) as shown in Ref.
\citep{wang2022relativistic}.

So what is the reason for the sign problem in Eq. (\ref{eq:hertz-2})?
The answer lies in the linear constitutive relations (\ref{eq:constitutive-lab})
defined in the lab frame. This is valid for a static medium and not
for a moving medium. The linear constitutive relations should be defined
in the medium's comoving frame as the relations for three-vector fields
and get modified in the lab frame in a nontrivial way \citep{Landau:1984,Rousseaux2013}.
The covariant form of the constitutive relations (\ref{eq:constitutive-relations})
meets this requirement and therefore leads to Eq. (\ref{eq:maxwell-move-medium})
having an implicit Lorentz covariance in the SVA.


\section{Integral forms of Faraday and Ampere laws for moving surfaces}

\label{sec:integral-form}The integral form of Maxwell equations can
be written in accordance with the differential form. However the integral
form involves the definition of the integrals over volumes, closed
or open surfaces and closed lines (loops). When these volumes, surfaces
and loops are moving in one specific frame, the integral form of the
equations in this frame becomes more subtle than expected. The subtlety
lies in the fact that these equations are in three-dimensional forms
instead of covariant forms. This is the case for Faraday and Ampere
laws which involve time derivatives of surface integrals as well as
loops integrals.

Let us first look at Faraday law in the following integral form in
the lab frame
\begin{equation}
\mathcal{E}_{EMF}=-\frac{1}{c}\frac{d\Phi(t)}{dt}=-\frac{1}{c}\frac{d}{dt}\int_{S}d\mathbf{S}\cdot\mathbf{B}(x),\label{eq:faraday}
\end{equation}
where $\mathcal{E}_{EMF}$ is the electromotive force and $\Phi(t)$
is the flux of magnetic field through a surface $S$.

When $S$ is static and fixed in the lab frame (not moving), there
is no ambiguity for $\mathcal{E}_{EMF}$ which is given by
\begin{equation}
\mathcal{E}_{EMF}=\varoint_{C}d\boldsymbol{l}\cdot\mathbf{E}(x),\label{eq:emf}
\end{equation}
where $C$ is the boundary of $S$. Because $S$ and $C$ are static
and fixed in the lab frame, the time derivative can be moved inside
the integral and work on $\mathbf{B}(x)=\mathbf{B}(t,\mathbf{x})$,
which gives the differential form of Faraday equation with the help
of Stokes theorem 
\begin{equation}
\boldsymbol{\nabla}\times\mathbf{E}(x)=-\frac{1}{c}\frac{\partial\mathbf{B}(x)}{\partial t}.\label{eq:faraday-lab}
\end{equation}

Now we consider the case that $S$ and $C$ are moving in the lab
frame with a low speed $v\ll c$. In this case we show the explicit
time dependence of the surface and its boundary as $S(t)$ and $C(t)$.
Then the time derivative of the flux in Eq. (\ref{eq:faraday}) becomes
\citep{Jackson:1998nia}
\begin{eqnarray}
\frac{d\Phi(t)}{dt} & = & \frac{1}{c}\int_{S(t)}d\mathbf{S}\cdot\frac{\partial\mathbf{B}(x)}{\partial t}+\frac{1}{c}\lim_{\Delta t\rightarrow0}\frac{1}{\Delta t}\left(\int_{S(t+\Delta t)}-\int_{S(t)}\right)d\mathbf{S}\cdot\mathbf{B}(x)\nonumber \\
 & = & \frac{1}{c}\int_{S(t)}d\mathbf{S}\cdot\frac{\partial\mathbf{B}(x)}{\partial t}-\frac{1}{c}\varoint_{C(t)}d\boldsymbol{l}\cdot[\mathbf{v}\times\mathbf{B}(x)],\label{eq:d-flux}
\end{eqnarray}
where the second term is from the change of $S(t)$. Using Faraday
equation in the lab frame, Eq. (\ref{eq:faraday-lab}), and then Stokes
theorem, we obtain 
\begin{eqnarray}
\frac{d\Phi(t)}{dt} & = & -\int_{S(t)}d\mathbf{S}\cdot\boldsymbol{\nabla}\times\mathbf{E}(x)-\frac{1}{c}\varoint_{C(t)}d\boldsymbol{l}\cdot[\mathbf{v}\times\mathbf{B}(x)]\nonumber \\
 & = & -\varoint_{C(t)}d\boldsymbol{l}\cdot\left[\mathbf{E}(x)+\frac{1}{c}\mathbf{v}\times\mathbf{B}(x)\right].
\end{eqnarray}
The above equation defines $\mathcal{E}_{EMF}$ for a moving $S(t)$
and $C(t)$ \citep{Jackson:1998nia}, 
\begin{equation}
\mathcal{E}_{EMF}=\varoint_{C(t)}d\boldsymbol{l}\cdot\left[\mathbf{E}(x)+\frac{1}{c}\mathbf{v}\times\mathbf{B}(x)\right].\label{eq:emf-3d}
\end{equation}
Obviously this is not the form in Eq. (\ref{eq:emf}) for the static
case. So Faraday equation in the integral form for a slowly moving
surface reads \citep{Jackson:1998nia}
\begin{equation}
\varoint_{C(t)}d\boldsymbol{l}\cdot\left[\mathbf{E}(x)+\frac{1}{c}\mathbf{v}\times\mathbf{B}(x)\right]=-\frac{1}{c}\frac{d}{dt}\int_{S(t)}d\mathbf{S}\cdot\mathbf{B}(x).\label{eq:faraday-move}
\end{equation}
Rewriting the term $\varoint_{C(t)}d\boldsymbol{l}\cdot(\mathbf{v}\times\mathbf{B})$
in Eq. (\ref{eq:d-flux}) into a surface integral using Stokes theorem,
Eq. (\ref{eq:faraday-move}) gives Faraday equation in the differential
form 
\begin{equation}
\boldsymbol{\nabla}\times\left[\mathbf{E}(x)+\frac{1}{c}\mathbf{v}\times\mathbf{B}(x)\right]=-\frac{1}{c}\left(\frac{\partial}{\partial t}+{\bf v}\cdot\boldsymbol{\nabla}\right)\mathbf{B}(x),
\end{equation}
which is just Eq. (\ref{eq:pauli-1}) given by Pauli and consistent
with Eq. (\ref{eq:faraday-lab}). This corresponds to case (d) in
Table \ref{tab:ME-form}. Note that the field in the loop integral
for the moving surface is the comoving field $\boldsymbol{\mathcal{E}}(x)=\mathbf{E}(x)+(1/c)\mathbf{v}\times\mathbf{B}$
instead of $\mathbf{E}(x)$. This is due to the fact that $\mathcal{E}_{EMF}$
measures the electromotive force in the moving loop $C(t)$, which
should include the Lorentz force $(1/c)\mathbf{v}\times\mathbf{B}$.

The integral form of Ampere law (equation) for the slowing moving
surface in the lab frame can be presented in a similar way. The resulting
equation reads 
\begin{eqnarray}
 &  & \varoint_{C(t)}d\boldsymbol{l}\cdot\left[{\bf H}(x)-\frac{{\bf v}}{c}\times{\bf D}(x)\right]\nonumber \\
 & = & \frac{1}{c}\int_{S(t)}d\mathbf{S}\cdot\left[\mathbf{J}_{f}(x)-\rho_{f}(x){\bf v}\right]+\frac{1}{c}\frac{d}{dt}\int_{S(t)}d\mathbf{S}\cdot\mathbf{D}(x),\label{eq:ampere-move}
\end{eqnarray}
which gives Ampere equation in the differential form
\begin{eqnarray}
 &  & \boldsymbol{\nabla}\times\left[{\bf H}(x)-\frac{{\bf v}}{c}\times{\bf D}(x)\right]\nonumber \\
 & = & \frac{1}{c}\left[\mathbf{J}_{f}(x)-\rho_{f}(x){\bf v}\right]+\frac{1}{c}\left(\frac{\partial}{\partial t}+{\bf v}\cdot\boldsymbol{\nabla}\right)\mathbf{D}(x),
\end{eqnarray}
which is just Eq. (\ref{eq:pauli-2}) given by Pauli and consistent
with the last line of Eq. (\ref{eq:three-dim-maxwell-1}). This corresponds
to case (d) in Table \ref{tab:ME-form}.

The integral and differential forms of Faraday and Ampere laws for
moving surfaces are summarized in Table \ref{tab:integral-form}.

\begin{table}
\caption{\label{tab:integral-form}The integral and differential forms of Faraday
and Ampere laws for the moving surface $S(t)$ with the boundary $C(t)$.
They are all consistent with Maxwell equations in the lab frame (and
in any frame of course).}
\global\long\def\arraystretch{1.8}

\begin{tabular}{c|c}
\hline 
Form & Faraday law\tabularnewline
\hline 
Integral & $\varoint_{C(t)}d\boldsymbol{l}\cdot\left[\mathbf{E}(x)+\frac{1}{c}\mathbf{v}\times\mathbf{B}(x)\right]=-\frac{1}{c}\frac{d}{dt}\int_{S(t)}d\mathbf{S}\cdot\mathbf{B}(x)$\tabularnewline
\hline 
Differential & $\boldsymbol{\nabla}\times\left[\mathbf{E}(x)+\frac{1}{c}\mathbf{v}\times\mathbf{B}(x)\right]=-\frac{1}{c}\left(\frac{\partial}{\partial t}+{\bf v}\cdot\boldsymbol{\nabla}\right)\mathbf{B}(x)$\tabularnewline
\hline 
 & Ampere law\tabularnewline
\hline 
Integral & $\varoint_{C(t)}d\boldsymbol{l}\cdot\left[{\bf H}(x)-\frac{{\bf v}}{c}\times{\bf D}(x)\right]=\frac{1}{c}\int_{S(t)}d\mathbf{S}\cdot\left[\mathbf{J}_{f}(x)-\rho_{f}(x){\bf v}\right]+\frac{1}{c}\frac{d}{dt}\int_{S(t)}d\mathbf{S}\cdot\mathbf{D}(x)$\tabularnewline
\hline 
Differential & $\boldsymbol{\nabla}\times\left[{\bf H}(x)-\frac{{\bf v}}{c}\times{\bf D}(x)\right]=\frac{1}{c}\left[\mathbf{J}_{f}(x)-\rho_{f}(x){\bf v}\right]+\frac{1}{c}\left(\frac{\partial}{\partial t}+{\bf v}\cdot\boldsymbol{\nabla}\right)\mathbf{D}(x)$\tabularnewline
\hline 
\end{tabular}
\end{table}

To ultimately remove such a subtlety, we should derive Faraday equation
in the covariant integral form \citep{MARX1975353}. Before we do
so, we have to define an arbitrary open surface $S$ and its boundary
(a closed curve) $C$ in Minkowski space. The world line of all points
$x^{\mu}$ on the curve forms a two-dimensional tube in Minkowski
space, which can be parameterized by two parameters. We choose a frame
four-vector $n^{\mu}$ which satisfied $n_{\mu}n^{\mu}=1$ and define
the proper time $\tau$ as 
\begin{equation}
n\cdot x\equiv n^{\mu}x_{\mu}=c\tau,
\end{equation}
The open surface $S$ can be parameterized by $x^{\mu}(\tau,w_{1},w_{2})$
at fixed $\tau$. Its boundary $C$ can be obtained by setting $w_{1}(\tau,\theta)$
and $w_{2}(\tau,\theta)$. We can define the total time derivative
of the magnetic flux in the covariant form 
\begin{eqnarray}
\frac{1}{c}\frac{d\Phi}{d\tau} & = & \frac{1}{c}\left[\int_{S(\tau)}d\sigma_{\mu\nu}\frac{\partial\widetilde{F}^{\mu\nu}}{\partial\tau}\right.\nonumber \\
 &  & \left.+\lim_{\Delta\tau\rightarrow0}\frac{1}{\Delta\tau}\left(\int_{C(\tau+\Delta\tau)}-\int_{C(\tau)}\right)d\sigma_{\lambda\rho}\widetilde{F}^{\lambda\rho}\right],\label{eq:flux}
\end{eqnarray}
where the area element $d\sigma^{\mu\nu}$ on $S(\tau)$ is defined
as 
\begin{equation}
d\sigma_{\mu\nu}=\frac{1}{2}\epsilon_{\mu\nu\alpha\beta}\frac{\partial x^{\alpha}}{\partial w_{1}}\frac{\partial x^{\beta}}{\partial w_{2}}dw_{1}dw_{2},
\end{equation}
and the area element $d\sigma_{\lambda\rho}$ on the boundary $C(\tau)$
is defined as 
\begin{equation}
d\sigma_{\lambda\rho}=\frac{1}{2}\epsilon_{\lambda\rho\alpha\beta}\left(\frac{\partial x^{\alpha}}{\partial\tau}-cn^{\alpha}\right)\frac{\partial x^{\beta}}{\partial\theta}d\tau d\theta.\label{eq:d-sigma-c}
\end{equation}
Substituting (\ref{eq:d-sigma-c}) into the second term of (\ref{eq:flux})
and using 
\begin{equation}
\frac{1}{c}\frac{\partial\widetilde{F}^{\mu\nu}}{\partial\tau}=\frac{\partial\widetilde{F}^{\mu\nu}}{\partial x^{\lambda}}n^{\lambda},
\end{equation}
we obtain 
\begin{eqnarray}
\frac{1}{c}\frac{d\Phi}{d\tau} & = & \int_{S(\tau)}d\sigma_{\mu\nu}\frac{\partial\widetilde{F}^{\mu\nu}}{\partial x^{\lambda}}n^{\lambda}\nonumber \\
 &  & -\varoint_{C(\tau)}d\theta F_{\alpha\beta}\left(\frac{1}{c}\frac{\partial x^{\alpha}}{\partial\tau}-n^{\alpha}\right)\frac{\partial x^{\beta}}{\partial\theta}.\label{eq:flux-1}
\end{eqnarray}
One can prove with the first equation of (\ref{eq:homogen-maxwell})
\begin{equation}
\int_{S(\tau)}d\sigma_{\mu\nu}\frac{\partial\widetilde{F}^{\mu\nu}}{\partial x^{\lambda}}n^{\lambda}=-\varoint_{C(\tau)}d\theta F_{\alpha\beta}n^{\alpha}\frac{\partial x^{\beta}}{\partial\theta}.
\end{equation}
Using the above equation in Eq. (\ref{eq:flux-1}), only the first
term inside the parenthesis survives, so the electromotive force in
the covariant form is given by 
\begin{equation}
\mathcal{E}_{EMF}=-\frac{1}{c}\frac{d\Phi}{d\tau}=\frac{1}{c}\varoint_{C(\tau)}dl^{\beta}F_{\alpha\beta}\frac{\partial x^{\alpha}}{\partial\tau},
\end{equation}
where $dl^{\beta}=d\theta(\partial x^{\beta}/\partial\theta)$ is
the line element of $C(\tau)$. If we let $\partial x^{\alpha}/\partial\tau=cu^{\alpha}$
and use Eq. (\ref{eq:extract-eb}), the above equation becomes 
\begin{equation}
\mathcal{E}_{EMF}=-\varoint_{C(\tau)}dl^{\mu}\mathcal{E}_{\mu}.
\end{equation}
We see that $\mathcal{E}_{EMF}$ is a loop integral of the electric
field $\mathcal{E}^{\mu}$. For example, one can choose 
\begin{equation}
n^{\mu}=(1,\mathbf{0}),\;\;\frac{\partial x^{\alpha}}{\partial\tau}=cu^{\alpha}\approx c(1,\mathbf{v}/c),\;\;dl^{\beta}=(0,d\boldsymbol{l}),
\end{equation}
then one can verify that $\mathcal{E}_{EMF}$ recovers the three-dimensional
form in (\ref{eq:emf-3d}).

The most important message we would like to deliver in this section
is: the integral forms of Faraday and Ampere equations (\ref{eq:faraday-move})
and (\ref{eq:ampere-move}) for slowly moving surfaces are consistent
with Maxwell equations in (\ref{eq:three-dim-maxwell-1}). The fields
in loop integrals must be those in the comoving frame, $\boldsymbol{\mathcal{E}}(x)$
and $\boldsymbol{\mathcal{H}}(x)$, not $\mathbf{E}(x)$ and $\mathbf{H}(x)$,
otherwise the resulting equations would be inconsistent with Maxwell
equations and lead to contradiction.


\section{Summary}

We derive a set of Maxwell equations for slowly moving media from
the Lorentz transformation in the small velocity approximation (SVA).
Our derivation is based on the method of field decomposition widely
used in relativistic magnetohydrodynamics, in which the four-vectors
of electric and magnetic fields with Lorentz covariance can be defined.
We start from the covariant form of Maxwell equations to derive these
equations by taking an expansion in the medium velocity $v/c$ and
keeping terms up to $O(v/c)$. These ``deformed'' Maxwell equations
are written in space-time of the lab frame, which can recover the
conventional form of Maxwell equations if all fields and space-time
coordinates are written in the comoving frame of the medium.

The Lorentz transformation plays the key role to maintain the conformality
of Maxwell equations: the time and charge density must also change
when transforming to a different frame even in the SVA, not just the
position and current density as in the Galilean transformation. This
marks the essential difference of the Lorentz transformation from
the Galilean one.

The integral forms of Faraday and Ampere equations (\ref{eq:faraday-move})
and (\ref{eq:ampere-move}) for slowly moving surfaces are consistent
with Maxwell equations in (\ref{eq:three-dim-maxwell-1}). The fields
in loop integrals over moving surfaces must be those in the comoving
frame instead of those in the lab frame, otherwise the resulting equations
would be inconsistent with Maxwell equations and lead to contradiction.
We also present Faraday equation in the covariant integral form in
which the electromotive force can be defined as the four-dimensional
loop integral of the comoving electric field, a Lorentz scalar independent
of the observer's frame.

From the results of this paper, no evidence is found to support an
extension or modification of Maxwell equations. 

\textit{Acknowledgments}. We thank Hao Chen, Xi Dai, Tian-Jun Li,
Chun Liu, Wan-Dong Liu, Wei Sha, Fei Wang, Qing Wang, and Jin-Min
Yang for helpful discussions. Our special thanks go to Zhong-Lin Wang
for insightful discussions which deepen our understanding of this
topic and broaden our knowledge on the applicability of the study
to many other fields than TENGs. S.P. and Q.W. are supported in part
by National Natural Science Foundation of China (NSFC) under Grants
12135011, 11890713 (a subgrant of 11890710) and 12075235.


\appendix

\section{Derivation of 3-dimensional Maxwell equations from covariant ones}

\label{sec:derivation-dim-3}In this appendix, we derive Maxwell equations
in 3-dimensional form from the covariant ones in Eqs. (\ref{eq:homogen-maxwell})
and (\ref{eq:source-maxwell}). The $\nu=0$ component of Eq. (\ref{eq:homogen-maxwell})
reads
\begin{eqnarray}
0 & = & \frac{1}{2}\epsilon^{i0\alpha\beta}\partial_{i}F_{\alpha\beta}=-\frac{1}{2}\epsilon^{0ijk}\partial_{i}F_{jk}\nonumber \\
 & = & \boldsymbol{\nabla}\cdot\mathbf{B}\,,
\end{eqnarray}
where we have used $F^{ij}=F_{ij}=-\epsilon_{ijk}\mathbf{B}_{k}$.
The $\nu=i$ component of Eq. (\ref{eq:homogen-maxwell}) reads 
\begin{eqnarray}
0 & = & \frac{1}{2}\epsilon^{0i\alpha\beta}\partial_{0}F_{\alpha\beta}+\frac{1}{2}\epsilon^{ji\alpha\beta}\partial_{j}F_{\alpha\beta}\nonumber \\
 & = & -\left(\frac{1}{c}\frac{\partial\mathbf{B}}{\partial t}+\boldsymbol{\nabla}\times\mathbf{E}\right)_{i}\,,
\end{eqnarray}
where we have used $F^{0i}=-F_{0i}=-\mathbf{E}_{i}$. The above equation
leads to Faraday's law 
\begin{equation}
\boldsymbol{\nabla}\times\mathbf{E}=-\frac{1}{c}\frac{\partial\mathbf{B}}{\partial t}\,.
\end{equation}
The $\nu=0$ component of Eq. (\ref{eq:source-maxwell}) reads
\begin{equation}
\partial_{i}F^{i0}=\boldsymbol{\nabla}\cdot\mathbf{E}=\rho\,.
\end{equation}
The $\nu=i$ component of Eq. (\ref{eq:source-maxwell}) reads 
\begin{eqnarray}
\frac{1}{c}J^{i} & = & \partial_{0}F^{0i}+\partial_{j}F^{ji}\nonumber \\
 & = & \left(-\frac{1}{c}\frac{\partial\mathbf{E}}{\partial t}+\boldsymbol{\nabla}\times\mathbf{B}\right)_{i}\,,
\end{eqnarray}
which leads to Ampere's law 
\begin{equation}
\boldsymbol{\nabla}\times\mathbf{B}=\frac{1}{c}\mathbf{J}+\frac{1}{c}\frac{\partial\mathbf{E}}{\partial t}\,.
\end{equation}
Then the above equations are put together into Eq. (\ref{eq:three-dim-maxwell}).

\section{Equations for $\boldsymbol{\mathcal{E}}$ and $\boldsymbol{\mathcal{B}}$
in SVA}

\label{sec:derivation-maxwell-move}Substituting Eq. (\ref{eq:fmunu})
into Eq. (\ref{eq:source-maxwell}), we obtain 
\begin{eqnarray}
0 & = & \frac{1}{2}\epsilon^{\mu\nu\alpha\beta}\partial_{\mu}F_{\alpha\beta}\nonumber \\
 & = & \epsilon^{\mu\nu\alpha\beta}u_{\beta}\partial_{\mu}\mathcal{E}_{\alpha}-u\cdot\partial\mathcal{B}^{\nu}+u^{\nu}(\partial\cdot\mathcal{B})\,.\label{eq:homo-maxwell}
\end{eqnarray}
We can write $u\cdot\partial$ and $\partial\cdot\mathcal{B}$ explicitly
\begin{eqnarray}
u\cdot\partial & = & \frac{1}{c}\gamma\left(\frac{\partial}{\partial t}+\mathbf{v}\cdot\boldsymbol{\nabla}\right),\nonumber \\
\partial\cdot\mathcal{B} & = & \left(\boldsymbol{\nabla}+\frac{1}{c^{2}}\frac{\partial}{\partial t}\mathbf{v}\right)\cdot\boldsymbol{\mathcal{B}}.
\end{eqnarray}

In the SVA up to $O(v/c)$, the $\nu=0$ component of Eq. (\ref{eq:homo-maxwell})
gives 
\begin{eqnarray}
0 & = & \epsilon^{\mu0\alpha\beta}u_{\beta}\partial_{\mu}\mathcal{E}_{\alpha}-u\cdot\partial\mathcal{B}^{0}+u^{0}(\partial\cdot\mathcal{B})\nonumber \\
 & \approx & \left(\boldsymbol{\nabla}+\frac{\mathbf{v}}{c^{2}}\frac{\partial}{\partial t}\right)\cdot\boldsymbol{\mathcal{B}}\,,\label{eq:divb}
\end{eqnarray}
where we have neglected $O(v^{2}/c^{2})$ term.

In the SVA up to $O(v)$, the $\nu=i$ component of Eq. (\ref{eq:homo-maxwell})
gives 
\begin{eqnarray}
0 & = & \epsilon^{\mu i\alpha\beta}u_{\beta}\partial_{\mu}\mathcal{E}_{\alpha}-u\cdot\partial\mathcal{B}^{i}+u^{i}(\partial\cdot\mathcal{B})\nonumber \\
 & \approx & \gamma\left[-\frac{1}{c^{2}}\mathbf{v}\times\frac{\partial}{\partial t}\boldsymbol{\mathcal{E}}-\boldsymbol{\nabla}\times\boldsymbol{\mathcal{E}}-\frac{1}{c}\left(\frac{\partial}{\partial t}+\mathbf{v}\cdot\boldsymbol{\nabla}\right)\boldsymbol{\mathcal{B}}\right]_{i}\,,
\end{eqnarray}
which leads to 
\begin{equation}
\left(\boldsymbol{\nabla}+\frac{\mathbf{v}}{c^{2}}\frac{\partial}{\partial t}\right)\times\boldsymbol{\mathcal{E}}=-\frac{1}{c}\left(\frac{\partial}{\partial t}+\mathbf{v}\cdot\boldsymbol{\nabla}\right)\boldsymbol{\mathcal{B}}\,,\label{eq:faraday-cal}
\end{equation}
where we have used Eq. (\ref{eq:divb}).

From Eqs. (\ref{eq:source-maxwell}) and (\ref{eq:fmunu}) we obtain
\begin{eqnarray}
\partial_{\mu}F^{\mu\nu}(x) & = & \partial_{\mu}\left[\mathcal{E}^{\mu}(x)u^{\nu}-\mathcal{E}^{\nu}(x)u^{\mu}+\epsilon^{\mu\nu\rho\sigma}u_{\rho}\mathcal{B}_{\sigma}(x)\right]\nonumber \\
 & = & u^{\nu}\partial\cdot\mathcal{E}-u\cdot\partial\mathcal{E}^{\nu}+\epsilon^{\mu\nu\rho\sigma}u_{\rho}\partial_{\mu}\mathcal{B}_{\sigma}\nonumber \\
 & = & \frac{1}{c}J^{\nu}\,.\label{eq:me-source}
\end{eqnarray}

In the SVA up to $O(v/c)$, we obtain the $\nu=0$ component of Eq.
(\ref{eq:me-source}) as 
\begin{eqnarray}
\partial_{\mu}F^{\mu0}(x) & = & u^{0}\partial\cdot\mathcal{E}-u\cdot\partial\mathcal{E}^{0}+\epsilon^{\mu0\rho\sigma}u_{\rho}\partial_{\mu}\mathcal{B}_{\sigma}\nonumber \\
 & \approx & \gamma\left[\boldsymbol{\nabla}\cdot\boldsymbol{\mathcal{E}}+\frac{\mathbf{v}}{c}\cdot(\boldsymbol{\nabla}\times\boldsymbol{\mathcal{B}})\right]=\rho\,.
\end{eqnarray}
Using Eq. (\ref{eq:ampere-law-cal}) and neglecting $O(v^{2})$ terms,
we obtain 
\begin{equation}
\left(\boldsymbol{\nabla}+\frac{\mathbf{v}}{c^{2}}\frac{\partial}{\partial t}\right)\cdot\boldsymbol{\mathcal{E}}=\rho-\frac{1}{c^{2}}\mathbf{v}\cdot\mathbf{J}\,.\label{eq:coulomb-cal}
\end{equation}

In the SVA up to $O(v/c)$, the $\nu=i$ component of Eq. (\ref{eq:me-source})
is simplified as 
\begin{eqnarray}
\frac{1}{c}J^{i} & = & \partial_{\mu}F^{\mu i}(x)\nonumber \\
 & = & u^{i}\partial\cdot\mathcal{E}-u\cdot\partial\mathcal{E}^{i}+\epsilon^{\mu i\rho\sigma}u_{\rho}\partial_{\mu}\mathcal{B}_{\sigma}\nonumber \\
 & \approx & \gamma\left[\frac{1}{c}\rho\mathbf{v}-\frac{1}{c}\left(\frac{\partial}{\partial t}+\mathbf{v}\cdot\boldsymbol{\nabla}\right)\boldsymbol{\mathcal{E}}+\frac{1}{c^{2}}\frac{\partial}{\partial t}(\mathbf{v}\times\boldsymbol{\mathcal{B}})+\boldsymbol{\nabla}\times\boldsymbol{\mathcal{B}}\right]_{i}\,,
\end{eqnarray}
which leads to 
\begin{eqnarray}
\left(\boldsymbol{\nabla}+\frac{\mathbf{v}}{c^{2}}\frac{\partial}{\partial t}\right)\times\boldsymbol{\mathcal{B}} & = & \frac{1}{c}(\mathbf{J}-\rho\mathbf{v})+\frac{1}{c}\left(\frac{\partial}{\partial t}+\mathbf{v}\cdot\nabla\right)\boldsymbol{\mathcal{E}}\,,\label{eq:ampere-law-cal}
\end{eqnarray}
where we have used Eq. (\ref{eq:coulomb-cal}).

Equations (\ref{eq:divb}), (\ref{eq:faraday-cal}), (\ref{eq:coulomb-cal})
and (\ref{eq:ampere-law-cal}) are Maxwell equations in moving frame
and put together into Eq. (\ref{eq:maxwell-move}).


\bibliographystyle{h-physrev}
\bibliography{maxwell-ref}

\end{document}